%% file: nonlinear.tex
\documentclass[12pt, onecolumn, draftcls]{IEEEtran}
\usepackage{graphics, epsfig, amsmath, amsfonts, amssymb, eucal}
\usepackage[nolists, nomarkers]{endfloat}
\newtheorem{prop}{Proposition}

\renewcommand{\baselinestretch}{1.5}

\input{rv_defs1}

\begin{document}

\title{A General Framework for Transmission with Transceiver Distortion and Some Applications}
\author{Wenyi Zhang, {\it Member, IEEE}
\thanks{W. Zhang is with Department of Electronic Engineering and Information Science, University of Science and Technology of China, Hefei, China. (e-mail: \texttt{wenyizha@ustc.edu.cn})}
\thanks{This research has been funded in part by National Science Foundation of China through grant 61071095, Research Fund for the Doctoral Program of Higher Education of China through grant 20103402120023, and the Fundamental Research Funds for the Central Universities of China.}
}
\maketitle
\thispagestyle{empty}

\begin{abstract}
A general theoretical framework is presented for analyzing information transmission over Gaussian channels with memoryless transceiver distortion, which encompasses various nonlinear distortion models including transmit-side clipping, receive-side analog-to-digital conversion, and others. The framework is based on the so-called generalized mutual information (GMI), and the analysis in particular benefits from the setup of Gaussian codebook ensemble and nearest-neighbor decoding, for which it is established that the GMI takes a general form analogous to the channel capacity of undistorted Gaussian channels, with a reduced ``effective'' signal-to-noise ratio (SNR) that depends on the nominal SNR and the distortion model. When applied to specific distortion models, an array of results of engineering relevance is obtained. For channels with transmit-side distortion only, it is shown that a conventional approach, which treats the distorted signal as the sum of the original signal part and a uncorrelated distortion part, achieves the GMI. For channels with output quantization, closed-form expressions are obtained for the effective SNR and the GMI, and related optimization problems are formulated and solved for quantizer design. Finally, super-Nyquist sampling is analyzed within the general framework, and it is shown that sampling beyond the Nyquist rate increases the GMI for all SNR. For example, with a binary symmetric output quantization, information rates exceeding one bit per channel use are achievable by sampling the output at four times the Nyquist rate.
\end{abstract}

\newpage
\setcounter{page}{1}

\section{Introduction}
\label{sec:introduction}

In digital communication systems, various forms of distortion are ubiquitous, acting as the main limiting factor for information transmission. Those distortions that come with the propagation of signal, such as shadowing and multipath fading, have received extensive research since the earliest era of digital communications \cite{biglieri98:it}. The current paper, alternatively, concerns with the other category of distortions that come mainly with the engineering of transceivers. This category of distortions encompasses a number of models of practical importance, including the clipping or saturation of transmitted waveforms due to power amplifier nonlinearity, the analog-to-digital conversion ({\it i.e.}, quantization) of received samples, and others. Such distortions are difficult to eliminate, and indeed people may deliberately introduce them, for practical reasons like hardware cost reduction and energy efficiency improvement.

We can usually approximate the aforementioned transceiver distortions as memoryless deterministic functions. Those functions, however, are generally nonlinear operations and thus break down the linearity in Gaussian channels. From a pure information-theoretic perspective, nonlinearity may not impose fundamental difficulty to our conceptual understanding, since the channel capacity is still the maximum of mutual information between the channel input and the distorted channel output. From an engineering perspective, however, the general mutual information maximization problem is usually less satisfactory in generating insights, especially when such maximization problems are analytically difficult, or even intractable, for general nonlinear channel models.

There are a number of existing works that seek to characterize the information-theoretic behavior of nonlinear transceiver distortion, largely scattered in the literature. In \cite{ochiai02:com}, the authors examined the channel capacity of clipped orthogonal frequency-division multiplexing (OFDM) systems, with the key approximation that the distortion due to clipping acts as an additional Gaussian noise. Such an approximation originates from a theorem due to Bussgang \cite{bussgang52:rle}, which implies that the output process of a Gaussian input process through a memoryless distortion device is the sum of a scaled input process and a distortion process which is uncorrelated with the input process. Regarding Nyquist-sampled real Gaussian channels with output quantization, an earlier study \cite{viterbi79:book} examined the achievable mutual information as the signal-to-noise ratio (SNR) decreases toward zero. Specifically, the numerical study therein revealed that for a binary symmetric output quantizer, the ratio between the capacity per channel use (c.u.) and the SNR approaches $1/\pi$, and that for a uniform octal ({\it i.e.}, $8$-level) output quantizer, this ratio is no less than $0.475$. In \cite{singh09:com}, the authors further established some general results for Nyquist-sampled real Gaussian channels, asserting that with a $K$-level output quantization, the capacity is achieved by choosing no more than $(K + 1)$ input levels, and that with a binary symmetric output quantization the capacity is indeed achieved by using a binary symmetric input distribution. For $K > 2$, however, it is necessary to use intensive numerical methods like the cutting-plane algorithm to compute the capacity. The authors of \cite{mezghani08:isit} addressed the capacity of multiple-input-multiple-output block-Rayleigh fading channels with binary symmetric output quantization. In \cite{koch10:arxiv}, the authors went beyond the Nyquist-sampled channel model, demonstrating that the low-SNR capacity of a real Gaussian channel with binary symmetric output quantization, when sampled at twice the Nyquist rate, is higher than that when sampled at the Nyquist rate. In \cite{koch11:arxiv}, the authors proved that by using a binary asymmetric output quantizer design, it is possible to achieve the low-SNR asymptotic capacity without output quantization.

Recognizing the challenge in working with channel capacity directly, we take an alternative route that seeks to characterize achievable information rates for certain specific encoding/decoding scheme. As the starting point of our study, in the current paper we consider a real Gaussian channel with general transceiver distortion, and focus on the Gaussian codebook ensemble and the nearest-neighbor decoding rule. We use the so-called generalized mutual information (GMI) \cite{ganti00:it, lapidoth02:it} to characterize the achievable information rate. As a performance measure for mismatched decoding, GMI has proved convenient and useful in several other scenarios such as multipath fading channels \cite{lapidoth02:it}. Herein, in our exercise with GMI, we aim at providing key engineering insights into the understanding and design of transceivers with nonlinearity. The nature of our approach is somewhat similar to that of \cite{salz95:com}, where the authors addressed the decoder design with a finite resolution constraint, using a performance metric akin to cutoff rate that also derives from a random-coding argument.

The motivation for using the performance measure of GMI and the Gaussian codebook ensemble coupled with the nearest-neighbor decoding is two-fold. On one hand, such an approach enables us to obtain an array of analytical results that are both convenient and insightful, and bears an ``operational'' meaning in that the resulting GMI is achievable, by the specific encoding/decoding scheme whose implementation does not heavily depend on the nonlinear distortion model. On the other hand, Gaussian codebook ensemble is a reasonable model for approximating the transmitted signals in many modern communication systems, in particular, those that employ higher-order modulation or multicarrier techniques like OFDM\footnote{In the current paper we confine ourselves to the single-carrier real Gaussian channel model, and will treat multicarrier transmission with nonlinear distortion in a separate work.}; and the nearest-neighbor decoding rule is also a frequently encountered solution in practice which is usually easier to implement than maximum-likelihood decoding, for channels with nonlinear characteristics. Nevertheless, we need to keep in mind that compared with capacity, the performance loss of GMI due to the inherently suboptimal encoding/decoding scheme used may not be negligible.

The central result in the current paper is a GMI formula, taking the form of $(1/2) \log (1 + \mathrm{SNR}_e)$, for real Gaussian channels\footnote{For complex Gaussian channels we also have an analogous result; see Supplementary Material \ref{appendix:basic-model-complex}.} with general transceiver distortion. Here $\mathrm{SNR}_e$ depends on the nominal SNR and the transceiver nonlinearity, and we may interpret it as the ``effective SNR'', due to its apparent similarity with the role of SNR in the capacity formula for Gaussian channels without distortion. The parameter $\mathrm{SNR}_e$ thus serves as a single-valued performance indicator, based on which we can, in a unified fashion, analyze the behavior of given transceivers, compare different distortion models, and optimize transceiver design.

Applying the aforementioned general GMI formula to specific distortion models, we obtain an array of results that are of engineering relevance. First, when the nonlinear distortion occurs at the transmitter only, we show that the Bussgang decomposition, which represents a received signal as the sum of a scaled input signal part and a distortion part which is uncorrelated with the input signal, is consistent with the GMI-maximizing nearest-neighbor decoding rule. This result validates the Gaussian clipping noise approximation for transmit-side clipping, as followed by the authors of \cite{ochiai02:com}.

Second, we evaluate the GMI for Nyquist-sampled channels with output quantization. For binary symmetric quantization, we find that the low-SNR asymptotic GMI coincides with the channel capacity. This observation is somewhat surprising, since the GMI is with respect to a suboptimal input distribution, namely the Gaussian codebook ensemble. On the other hand, there exists a gap between high-SNR asymptotic GMI and the channel capacity, revealing the penalty of suboptimal input distribution when the effect of noise is negligible. For symmetric quantizers with more than two quantization levels, we formulate a quantizer optimization problem that yields the maximum GMI, and present numerical results for uniform and optimized quantizers. As an example of our results, we show that for octal quantizers, the low-SNR asymptotic GMI is higher than the known lower bound of channel capacity in the literature \cite{viterbi79:book}.

Finally, we explore the benefit of super-Nyquist sampling. Considering a real Gaussian channel with a bandlimited pulse-shaping function and with general memoryless output distortion, we obtain a formula for its GMI, when the channel output is uniformly sampled at $L$ times the Nyquist rate. We then particularize to the case of binary symmetric output quantization. We demonstrate through numerical evaluation that super-Nyquist sampling leads to benefit in terms of increased GMI over all SNR, for different values of $L$. In the low-SNR regime, the asymptotic GMI we obtain for $L = 2$ with a carefully chosen pulse-shaping function almost coincides with the known lower bound of channel capacity in the literature \cite{koch10:arxiv}. In the high-SNR regime, we make an interesting observation that, when the sampling rate is sufficiently high, the GMI becomes greater than one bit/c.u.. At first glance, this result is surprising since the output quantization is binary; however, it is in fact reasonable, because for each channel input symbol, there are multiple binary output symbols due to super-Nyquist sampling, and the amount of information carried by the Gaussian codebook ensemble exceeds one bit per input symbol.

We organize the remaining part of the paper as follows. Section \ref{sec:basic-model-real} describes the general Nyquist-sampled channel model and establishes the general GMI formula. Section \ref{sec:input-distortion} treats the scenario where only transmit-side distortion exists, revisiting the well-known decomposition of Bussgang's theorem. Section \ref{sec:1-bit} treats the channel model with binary symmetric output quantization. Section \ref{sec:multi-bit} treats symmetric output quantizers with more than two quantization levels. Section \ref{sec:super-nyquist} explores the benefit of super-Nyquist sampling. Finally Section \ref{sec:conclusions} concludes the paper. Auxiliary technical derivations and other supporting results are archived in the Supplementary Material.

\section{General Framework for Real-Valued Nyquist-Sampled Channels}
\label{sec:basic-model-real}

\begin{figure}[h]
\epsfxsize=2in
\epsfclipon
\centerline{\epsffile{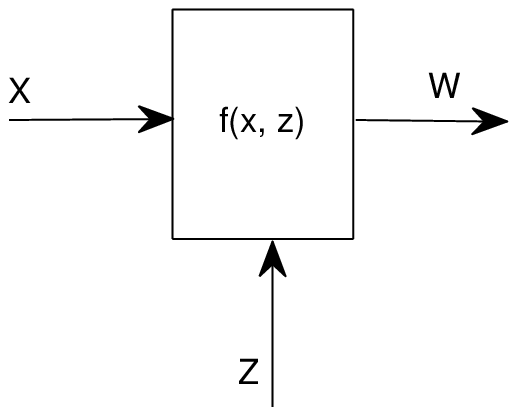}}
\caption{Illustration of the general channel model with distortion.}
\label{fig:model}
\end{figure}

With Nyquist sampling, it loses no generality to consider a discrete-time channel model, with a sequence $\{\rvz_k\}$ of independent and identically distributed (i.i.d.) real Gaussian noise, {\it i.e.}, $\rvz_\cdot \sim \mathcal{N}(0, \sigma^2)$. The channel input symbols constitute a sequence $\{\rvx_k\}$. Without distortion, the received signal is $\rvy_\cdot = \rvx_\cdot + \rvz_\cdot$. However, the distortion may affect both the channel input and the channel output. A memoryless distortion, in general form, is a deterministic mapping $f(\cdot)$, which transforms a pair of channel input symbol and noise sample $(x, z)$ into a real number $f(x, z)$. Hence the channel observation at the decoder is
\begin{eqnarray}
\label{eqn:channel-distorted}
\rvw_k = f(\rvx_k, \rvz_k), \quad \mbox{for}\; k = 1, 2, \ldots, n,
\end{eqnarray}
where $n$ denotes the codeword length; see the illustration in Figure \ref{fig:model}. We note that, such a form of distortion can describe the case where the channel output $\rvy_\cdot = \rvx_\cdot + \rvz_\cdot$ is distorted, {\it i.e.}, $w = f(x, z) = f_o(x + z)$, or the case where the channel input $\rvx_\cdot$ is distorted by the transmitter, {\it i.e.}, $w = f(x, z) = f_i(x) + z$, or the case where both input and output are distorted, {\it i.e.}, $w = f(x, z) = f_o(f_i(x) + z)$.

For transmission, the source selects a message $\rvm$ from $\mathcal{M} = \{1, 2, \ldots, \left\lfloor e^{nR} \right\rfloor\}$ uniformly randomly, and maps the selected message to a transmitted codeword, which is a length-$n$ real sequence, $\{\rvx_k(\rvm)\}_{k = 1}^n$. We restrict the codebook to be an i.i.d. $\mathcal{N}(0, \mathcal{E}_s)$ ensemble. That is, each codeword is a sequence of $n$ i.i.d. $\mathcal{N}(0, \mathcal{E}_s)$ random variables, and all the codewords are mutually independent. Such a choice of codebook ensemble satisfies the average power constraint $\frac{1}{n} \sum_{k = 1}^n \mathbf{E} \rvx_k^2(\rvm) \leq \mathcal{E}_s$. We thus define the nominal SNR as $\mathrm{SNR} = \mathcal{E}_s/\sigma^2$.

As is well known, when transceiver distortion is absent ({\it i.e.}, $w = y$), as the codeword length $n$ grows without bound, the Gaussian codebook ensemble achieves the capacity of the channel, $\frac{1}{2}\log(1 + \mathrm{SNR})$. In the following, we proceed to investigate the GMI when the channel experiences the memoryless nonlinear distortion $f(\cdot)$.

To proceed, we restrict the decoder to follow a nearest-neighbor rule, which, upon observing $\{w_k\}_{k = 1}^n$, computes for all possible messages, the distance metric,
\begin{eqnarray}
\label{eqn:nndec}
D(m) = \frac{1}{n}\sum_{k = 1}^n \left[ w_k - a x_k(m)\right]^2, \quad m \in \mathcal{M},
\end{eqnarray}
and decides the received message as $\hat{m} = \mathrm{arg}\min_{m \in \mathcal{M}} D(m)$. In (\ref{eqn:nndec}), the parameter $a$ is selected appropriately for optimizing the decoding performance. We note that, the nearest-neighbor decoder (with $a = 1$) coincides with the optimal (maximum-likelihood) decoder in the absence of distortion, but is in general suboptimal (mismatched) for the distorted channel (\ref{eqn:channel-distorted}).

In the subsequent development in this section, we characterize an achievable rate which guarantees that the average probability of decoding error decreases to zero as $n \rightarrow \infty$, for Gaussian codebook ensemble and the nearest-neighbor decoding rule, following the argument used in \cite{lapidoth02:it}. When we consider the average probability of decoding error averaged over both the message set and the Gaussian codebook ensemble, due to the symmetry in the codebook, it suffices to condition upon the scenario where the message $m = 1$ is selected for transmission.

With $m = 1$, we have
\begin{eqnarray}
\lim_{n \rightarrow \infty} D(1) &=& \lim_{n \rightarrow \infty} \frac{1}{n} \sum_{k = 1}^n \left[ \rvw_k - a \rvx_k(1)\right]^2 = \lim_{n \rightarrow \infty} \frac{1}{n} \sum_{k = 1}^n \left[ f(\rvx_k, \rvz_k) - a \rvx_k(1)\right]^2\nonumber\\
&=& \mathbf{E}\left\{\left[f(\rvx, \rvz) - a\rvx\right]^2\right\} \quad \mbox{a.s.}
\end{eqnarray}
where $\rvx \sim \mathcal{N}(0, \mathcal{E}_s)$ and $\rvz \sim \mathcal{N}(0, \sigma^2)$, from the law of large numbers.

The exponent of the probability of decoding error is the GMI, given by
\begin{eqnarray}
I_{\mathrm{GMI}} = \sup_{\theta < 0} \left(\theta \mathbf{E}\left\{\left[f(\rvx, \rvz) - a\rvx\right]^2\right\} - \Lambda(\theta)\right),
\end{eqnarray}
where
\begin{eqnarray}
\Lambda(\theta) &=& \lim_{n \rightarrow \infty} \frac{1}{n} \Lambda_n(n \theta),\\
\label{eqn:Lambda_n}
\Lambda_n(n\theta) &=& \log \mathbf{E}\left\{\left.e^{n\theta D(m)}\right| \rvw_k, k = 1, \ldots, n \right\}, \quad \forall m \neq 1.
\end{eqnarray}
From Chernoff's bound and the union upper bounding technique, we see that as long as the information rate is less than $I_\mathrm{GMI}$, the average probability of decoding error decreases to zero as $n \rightarrow \infty$. Therefore, the GMI serves as a reasonable lower bound for the achievable information rate for a given codebook ensemble and a given decoding rule.

After the mathematical manipulation given in Supplementary Material \ref{appendix:proof-gmi-real}, we establish the following result.
\begin{prop}
\label{prop:gmi-real}
With Gaussian codebook ensemble and nearest-neighbor decoding, the GMI of the distorted channel (\ref{eqn:channel-distorted}) is
\begin{eqnarray}
\label{eqn:gmi-real}
I_\mathrm{GMI} = \frac{1}{2}\log \left(1 + \frac{\Delta}{1 - \Delta}\right),
\end{eqnarray}
where the parameter $\Delta$ is
\begin{eqnarray}
\Delta = \frac{\left\{ \mathbf{E}[ f(\rvx, \rvz) \rvx ]\right\}^2}{\mathcal{E}_s \mathbf{E}[f(\rvx, \rvz)]^2}.
\end{eqnarray}
The corresponding optimal choice of the decoding scaling parameter $a$ is $a_\mathrm{opt} = \mathbf{E}\left[ f(\rvx, \rvz) \rvx \right]/\mathcal{E}_s$.
\end{prop}

We readily see that $\Delta$ is the squared correlated coefficient between the channel input $\rvx$ and the distorted channel output $f(\rvx, \rvz)$, which is upper bounded by one, from Cauchy-Schwartz inequality. A larger value of $\Delta$ corresponds to a higher effective SNR.

When contrasted with the capacity of the undistorted channel, $\frac{1}{2}\log(1 + \mathrm{SNR})$, we can define the effective SNR of the distorted channel as $\mathrm{SNR}_e = \frac{\Delta}{1 - \Delta}$.

As an immediate verification, consider the undistorted channel $\rvw_\cdot = \rvx_\cdot + \rvz_\cdot$, for which we have $\Delta = {\mathcal{E}_s}/(\mathcal{E}_s + \sigma^2)$. Consequently, the effective SNR is $\mathrm{SNR}_e = \mathcal{E}_s/\sigma^2$, leading to the capacity of the undistorted channel.

It is perhaps worth noting that, the derivation of the GMI in fact does not require $\rvz_\cdot$ be Gaussian. Indeed, as long as $\{\rvz_k\}$ is an ergodic process and is independent of $\{\rvx_k\}$, the general result of Proposition \ref{prop:gmi-real} holds. However, for simplicity, in the current paper we confine ourselves to i.i.d. Gaussian noise, and do not pursue this issue further.

\subsubsection*{Remark on Antipodal Codebook Ensemble}

The foregoing analysis of GMI applies to any input distribution. Here, consider antipodal inputs, {\it i.e.}, $\rvx_k(m)$ takes $\sqrt{\mathcal{E}_s}$ and $-\sqrt{\mathcal{E}_s}$ with probability $1/2$, respectively. All the codeword symbols are mutually independent. Again, we consider a nearest-neighbor decoding rule, with distance metric in form of (\ref{eqn:nndec}). Following the same line of analysis as that for the Gaussian codebook ensemble, we have
\begin{eqnarray}
\label{eqn:gmi-antipodal}
I_\mathrm{GMI} = \sup_{t \in \mathbb{R}} \left(t \mathbf{E}[\rvx f(\rvx, \rvz)] - \mathbf{E}\log\cosh(t\sqrt{\mathcal{E}_s} f(\rvx, \rvz)) \right),
\end{eqnarray}
and the optimal value of $t$ should satisfy
\begin{eqnarray}
\label{eqn:gmi-antipodal-t}
\mathbf{E}\left[\sqrt{\mathcal{E}_s} f(\rvx, \rvz) \cdot \tanh(t \sqrt{\mathcal{E}_s} f(\rvx, \rvz))\right] = \mathbf{E}[\rvx f(\rvx, \rvz)].
\end{eqnarray}
Supplementary Material \ref{appendix:antipodal}. The evaluation of the GMI is usually more difficult than that for the Gaussian codebook ensemble.

\section{Channels with Transmit-Side Distortion: Bussgang Revisited}
\label{sec:input-distortion}

In this section, we briefly consider the scenario where only the channel input is distorted, {\it i.e.}, $w = f_i(x) + z$. Since $\rvx$ and $\rvz$ are independent, the optimal choice of the decoding scaling parameter becomes
\begin{eqnarray}
\label{eqn:scaling-input-distortion}
a_\mathrm{opt} = \frac{\mathbf{E}[(f_i(\rvx) + \rvz)\rvx]}{\mathcal{E}_s} = \frac{\mathbf{E} [\rvx f_i(\rvx)]}{\mathcal{E}_s}.
\end{eqnarray}
The resulting value of $\Delta$ is
\begin{eqnarray}
\Delta = \frac{\left\{\mathbf{E} [\rvx f_i(\rvx)]\right\}^2}{\mathcal{E}_s \left(\mathbf{E} [f_i(\rvx)]^2 + \sigma^2 \right)},
\end{eqnarray}
and the effective SNR is
\begin{eqnarray}
\mathrm{SNR}_e = \frac{\Delta}{1 - \Delta} = \frac{\left\{\mathbf{E} [\rvx f_i(\rvx)]\right\}^2}{\mathcal{E}_s \left(\mathbf{E} [f_i(\rvx)]^2 + \sigma^2 \right) - \left\{\mathbf{E} [\rvx f_i(\rvx)]\right\}^2}.
\end{eqnarray}

Inspecting $a_\mathrm{opt}$ in (\ref{eqn:scaling-input-distortion}), we notice that it leads to the following decomposition of $f_i(\rvx)$:
\begin{eqnarray}
f_i(\rvx) = a_\mathrm{opt} \rvx + \rvv,
\end{eqnarray}
where the distortion $\rvv$ is uncorrelated with the input $\rvx$. Recalling the Bussgang decomposition \cite{ochiai02:com}, we conclude that, when there is only transmit-side distortion, the optimal decoding scaling parameter in the nearest-neighbor decoding rule coincides with that suggested by Bussgang's theorem. Note that this conclusion does not hold in general when receive-side distortion exists.

\section{Channels with Binary Symmetric Output Quantization}
\label{sec:1-bit}

In this section, we consider the scenario where the channel output $\rvy = \rvx + \rvz$ passes through a binary symmetric hard-limiter to retain its sign information only. This is also called one-bit/mono-bit quantization/analog-to-digital conversion, and we can write it as $w = f(x, z) = \mathrm{sgn}(x + z)$.

For this scenario, we have
\begin{eqnarray}
\label{eqn:Delta-1-bit}
\Delta = \frac{\left\{\mathbf{E}[\rvx\cdot \mathrm{sgn}(\rvx + \rvz)]\right\}^2}{\mathcal{E}_s},
\end{eqnarray}
where we use the fact that the average output power $\mathbf{E}[\mathrm{sgn}(\rvx + \rvz)]^2$ is unity. Now in order to facilitate the evaluation of the expectation in the numerator in (\ref{eqn:Delta-1-bit}), we introduce the ``partial mean'' of the random variable $\rvx \sim \mathcal{N}(0, \mathcal{E}_s)$
\begin{eqnarray}
F(z) = \int_z^\infty \frac{x}{\sqrt{2\pi \mathcal{E}_s}} e^{-\frac{x^2}{2\mathcal{E}_s}} dx = \sqrt{\frac{\mathcal{E}_s}{2\pi}} \exp\left(-\frac{z^2}{2\mathcal{E}_s}\right),
\end{eqnarray}
which is an even function of $z \in (-\infty, \infty)$. We denote by $p_\rvx(x)$ and $p_\rvz(z)$ the probability density functions of $\rvx \sim \mathcal{N}(0, \mathcal{E}_s)$ and $\rvz \sim \mathcal{N}(0, \sigma^2)$, respectively, and proceed as
\begin{eqnarray}
\label{eqn:x-sgny}
&&\mathbf{E}[\rvx \cdot \mathrm{sgn}(\rvx + \rvz)] = \iint_{x + z > 0} x p_\rvx(x)p_\rvz(z) dx dz - \iint_{x + z < 0} x p_\rvx(x)p_\rvz(z) dx dz\nonumber\\
&=& 2\iint_{x + z > 0} x p_\rvx(x)p_\rvz(z) dx dz = 2\int_{-\infty}^\infty p_\rvz(z) F(-z) dz = \mathcal{E}_s \sqrt{\frac{2}{\pi (\mathcal{E}_s + \sigma^2)}}.
\end{eqnarray}
This leads to
\begin{eqnarray}
\Delta = \frac{\mathcal{E}_s^2 \frac{2}{\pi (\mathcal{E}_s + \sigma^2)}}{\mathcal{E}_s} = \frac{2\mathcal{E}_s}{\pi (\mathcal{E}_s + \sigma^2)},
\end{eqnarray}
and
\begin{eqnarray}
\mathrm{SNR}_e = \frac{\Delta}{1 - \Delta} = \frac{2\mathcal{E}_s}{(\pi - 2) \mathcal{E}_s + \pi \sigma^2}.
\end{eqnarray}
So we get the following asymptotic behavior:
\begin{itemize}
\item High SNR: When $\mathrm{SNR} = \mathcal{E}_s/\sigma^2 \rightarrow \infty$,
\begin{eqnarray}
\mathrm{SNR}_e &=& \frac{2}{\pi - 2} - \frac{2\pi}{(\pi - 2)^2}\frac{1}{\mathrm{SNR}} + o(\frac{1}{\mathrm{SNR}}),\\
I_\mathrm{GMI} &=& \frac{1}{2}\log \frac{\pi}{\pi - 2} - \frac{1}{\pi - 2} \frac{1}{\mathrm{SNR}} + o(\frac{1}{\mathrm{SNR}}).
\end{eqnarray}
\item Low SNR: When $\mathrm{SNR} \rightarrow 0$,
\begin{eqnarray}
\mathrm{SNR}_e &=& \frac{2}{\pi} \mathrm{SNR} - \frac{2(\pi - 2)}{\pi^2} \mathrm{SNR}^2 + o(\mathrm{SNR}^2),\\
I_\mathrm{GMI} &=& \frac{1}{\pi} \mathrm{SNR} - \frac{\pi - 1}{\pi^2} \mathrm{SNR}^2 + o(\mathrm{SNR}^2).
\end{eqnarray}
\end{itemize}

\begin{figure}[ht]
\centering
\includegraphics[width=3.5in]{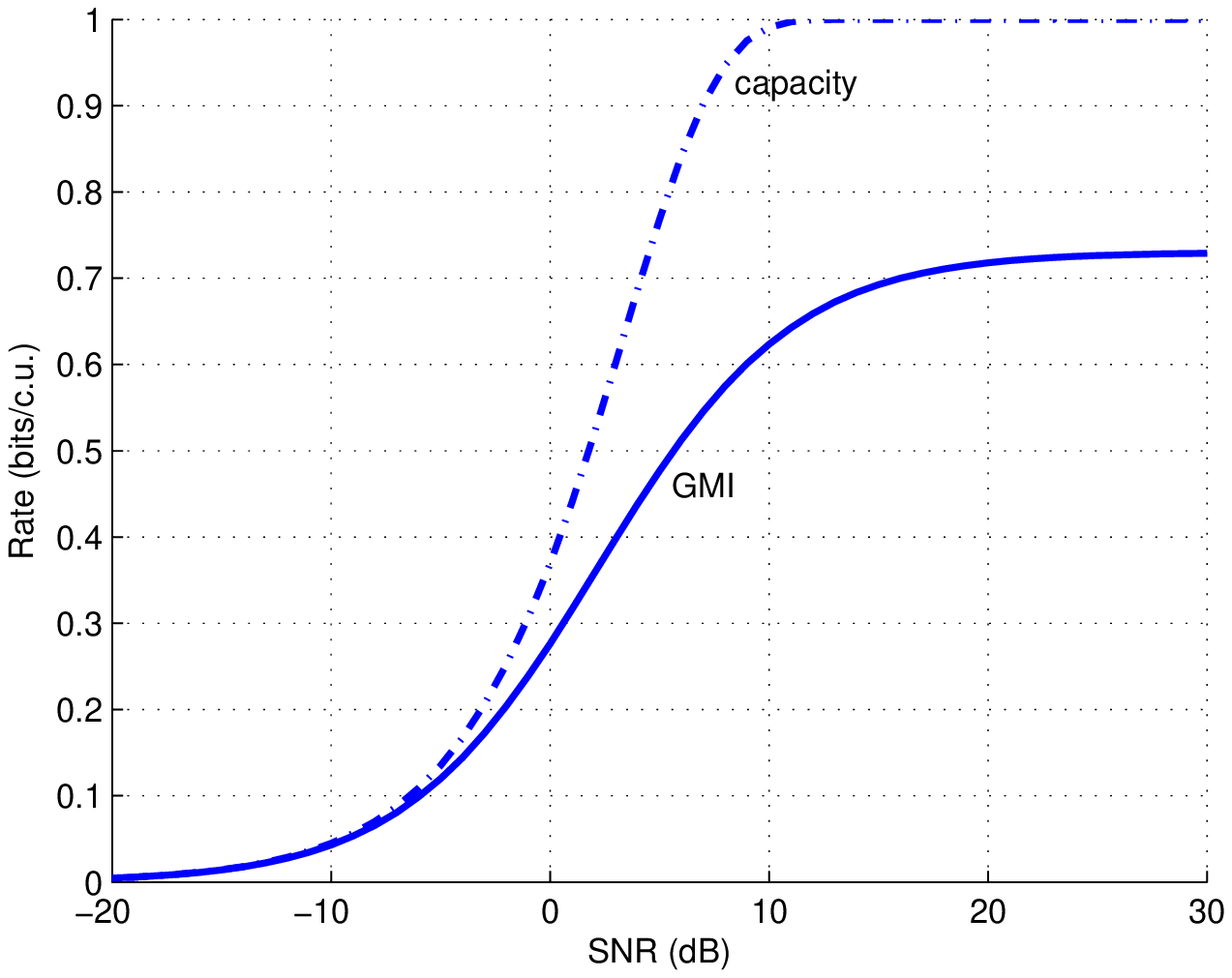}
\caption{The GMI and the channel capacity of the real Gaussian channel with binary symmetric output quantization.}
\label{fig:binary-adc-nyquist}
\end{figure}

We make two observations. First, at high SNR, the GMI converges to $0.7302$ bits/c.u., strictly less than the limit of the channel capacity, $1$ bit/c.u., revealing that the suboptimal Gaussian codebook ensemble leads to non-negligible penalty when the effect of distortion is dominant. Second, at low SNR, the ratio between the GMI and the SNR converges to $1/\pi$, and thus asymptotically coincides with the behavior of the channel capacity \cite{viterbi79:book}. Intuitively, this is because in the low-SNR regime the effect of noise is dominant, and thus the channel is approximately still Gaussian. In Figure \ref{fig:binary-adc-nyquist} we plot the GMI $I_\mathrm{GMI}$ and the channel capacity $C = 1 - H_2(Q(\sqrt{\mathcal{E}_s/\sigma^2}))$ \cite{singh09:com} versus $\mathrm{SNR}$. The different behaviors of the GMI in the two regimes are evident in the figure.

\section{Channels with Multi-Bit Output Quantization}
\label{sec:multi-bit}

In this section, we continue the exploration of output quantization and consider specifically the scenario where the channel output $\rvy$ passes through a $2M$-level symmetric quantizer, as
\begin{eqnarray}
w = f(x + z) = r_i \cdot \mathrm{sgn}(x + z) \quad \mbox{if}\; |x + z| \in [\alpha_{i - 1}, \alpha_i),
\end{eqnarray}
for $i = 1, \ldots, M$, where $\alpha_0 = 0 < \alpha_1 < \ldots < \alpha_M = \infty$. The parameters include the reconstruction points $\{r_1, \ldots, r_M\}$, and the quantization thresholds $\{\alpha_1, \ldots, \alpha_{M - 1}\}$. Note that with $2M$ levels, the quantizer bit-width is $(\log_2 M + 1)$ bits.

For a $2M$-level symmetric quantizer, we can evaluate that (see Supplementary Material \ref{appendix:2M-ADC-Delta-proof})
\begin{eqnarray}
\label{eqn:2M-ADC-Delta-1}
\mathbf{E}[f(\rvx + \rvz)]^2 = 2 \sum_{i = 1}^M r_i^2 \left[Q\left(\frac{\alpha_{i - 1}}{\sqrt{\mathcal{E}_s + \sigma^2}}\right) - Q\left(\frac{\alpha_i}{\sqrt{\mathcal{E}_s + \sigma^2}}\right)\right],
\end{eqnarray}
where the Q-function is $Q(z) = \frac{1}{\sqrt{2\pi}} \int_z^\infty e^{-x^2/2} dx$, and
\begin{eqnarray}
\label{eqn:2M-ADC-Delta-2}
\mathbf{E}[f(\rvx + \rvz) \rvx] = \mathcal{E}_s\sqrt{\frac{2}{\pi(\mathcal{E}_s + \sigma^2)}}\sum_{i = 1}^M r_i \left[e^{-\frac{\alpha_{i - 1}^2}{2(\mathcal{E}_s + \sigma^2)}} - e^{-\frac{\alpha_i^2}{2(\mathcal{E}_s + \sigma^2)}}\right].
\end{eqnarray}
To further simplify the notation, define $\tilde{Q}(z) = \frac{1}{2\sqrt{\pi}} \int_0^z (-\log x)^{-1/2} dx$ for $z \in [0, 1)$,\footnote{We have $\tilde{Q}(z) = Q(\sqrt{-2 \log z}) = (1/2)\cdot \mathrm{erfc}(\sqrt{-\log z})$.} and introduce $t_i = e^{-\frac{\alpha_i^2}{2(\mathcal{E}_s + \sigma^2)}}$ for $i = 0, 1, \ldots, M$ with $t_0 = 1 > t_1 > \ldots > t_M = 0$. We thus can rewrite
\begin{eqnarray}
\mathbf{E}[f(\rvx + \rvz)]^2 &=& 2\sum_{i = 1}^M r_i^2 [\tilde{Q}(t_{i - 1}) - \tilde{Q}(t_i)],\\
\mathbf{E}[f(\rvx + \rvz) \rvx] &=& \mathcal{E}_s\sqrt{\frac{2}{\pi(\mathcal{E}_s + \sigma^2)}} \sum_{i = 1}^M r_i (t_{i - 1} - t_i).
\end{eqnarray}
These lead to
\begin{eqnarray}
\label{eqn:delta-multi-bit}
\Delta = \frac{\mathcal{E}_s}{\pi (\mathcal{E}_s + \sigma^2)} \frac{\left[\sum_{i = 1}^M r_i (t_{i - 1} - t_i) \right]^2}{\sum_{i = 1}^M r_i^2 [\tilde{Q}(t_{i - 1}) - \tilde{Q}(t_i)]}.
\end{eqnarray}
In (\ref{eqn:delta-multi-bit}), the second term is independent of the SNR, and can be optimized separately. Let us denote this term by $K_{\underline{r}, \underline{t}}$, and write $\Delta = \frac{\mathcal{E}_s K_{\underline{r}, \underline{t}}}{\pi (\mathcal{E}_s + \sigma^2)}$. We consequently have the following effective SNR:
\begin{eqnarray}
\label{eqn:eff-snr-mbit}
\mathrm{SNR}_e = \frac{K_{\underline{r}, \underline{t}} \mathcal{E}_s}{(\pi - K_{\underline{r}, \underline{t}}) \mathcal{E}_s + \pi \sigma^2}.
\end{eqnarray}
\begin{itemize}
\item High SNR: When $\mathrm{SNR} \rightarrow \infty$,
\begin{eqnarray}
\mathrm{SNR}_e &=& \frac{K_{\underline{r}, \underline{t}}}{\pi - K_{\underline{r}, \underline{t}}} - \frac{K_{\underline{r}, \underline{t}}\pi}{(\pi - K_{\underline{r}, \underline{t}})^2}\frac{1}{\mathrm{SNR}} + o(\frac{1}{\mathrm{SNR}}),\\
I_\mathrm{GMI} &=& \frac{1}{2}\log \frac{\pi}{\pi - K_{\underline{r}, \underline{t}}} - \frac{K_{\underline{r}, \underline{t}}}{2(\pi - K_{\underline{r}, \underline{t}})} \frac{1}{\mathrm{SNR}} + o(\frac{1}{\mathrm{SNR}}).
\end{eqnarray}
\item Low SNR: When $\mathrm{SNR} \rightarrow 0$,
\begin{eqnarray}
\mathrm{SNR}_e &=& \frac{K_{\underline{r}, \underline{t}}}{\pi} \mathrm{SNR} - \frac{K_{\underline{r}, \underline{t}}(\pi - K_{\underline{r}, \underline{t}})}{\pi^2} \mathrm{SNR}^2 + o(\mathrm{SNR}^2),\\
I_\mathrm{GMI} &=& \frac{K_{\underline{r}, \underline{t}}}{2\pi} \mathrm{SNR} - \frac{K_{\underline{r}, \underline{t}}(\pi - K_{\underline{r}, \underline{t}}/2)}{2\pi^2} \mathrm{SNR}^2 + o(\mathrm{SNR}^2).
\end{eqnarray}
\end{itemize}

It is thus apparent that the value of $K_{\underline{r}, \underline{t}}$ determines the system performance, for all SNR. We hence seek to maximize
\begin{eqnarray}
K_{\underline{r}, \underline{t}} = \frac{\left[\sum_{i = 1}^M r_i (t_{i - 1} - t_i) \right]^2}{\sum_{i = 1}^M r_i^2 [\tilde{Q}(t_{i - 1}) - \tilde{Q}(t_i)]},
\end{eqnarray}
where $t_0 = 1 > t_1 > \ldots > t_M = 0$ and $r_i \geq 0$ for all $i = 1, \ldots, M$.

Taking the partial derivatives of $K_{\underline{r}, \underline{t}}$ with respect to $r_i$, $i = 1, \ldots, M$, and enforcing them to vanish, we have that the following set of equations needs to hold for maximizing $K_{\underline{r}, \underline{t}}$,
\begin{eqnarray}
r_i = \frac{t_{i - 1} - t_i}{\tilde{Q}(t_{i - 1}) - \tilde{Q}(t_i)} \frac{\sum_{j = 1}^M r_j^2 [\tilde{Q}(t_{j - 1}) - \tilde{Q}(t_j)]}{\sum_{j = 1}^M r_j (t_{j - 1} - t_j)}, \quad i = 1, \ldots, M.
\end{eqnarray}
Substituting these $\{r_i\}$ into $K_{\underline{r}, \underline{t}}$ and simplifying the resulting expression, we obtain
\begin{eqnarray}
K_{\underline{t}} = \max_{\underline{r}} K_{\underline{r}, \underline{t}} = \sum_{i = 1}^M \frac{(t_{i - 1} - t_i)^2}{\tilde{Q}(t_{i - 1}) - \tilde{Q}(t_i)}.
\end{eqnarray}
That is, the optimal quantizer design should solve the following maximization problem:
\begin{eqnarray}
\label{eqn:maximization-Kt}
\max_{\underline{t}} \sum_{i = 1}^M \frac{(t_{i - 1} - t_i)^2}{\tilde{Q}(t_{i - 1}) - \tilde{Q}(t_i)},\quad \mbox{s.t.}\quad t_0 = 1 > t_1 > \ldots > t_M = 0.
\end{eqnarray}

{\it Example:} Fine quantization, $\max_{i = 1, \ldots, M} (t_{i - 1} - t_i) \rightarrow 0$

In this case, the following approximation becomes accurate:
\begin{eqnarray}
\frac{\tilde{Q}(t_{i - 1}) - \tilde{Q}(t_i)}{t_{i - 1} - t_i} \approx \tilde{Q}^\prime (t_{i - 1}), \quad \forall i = 1, \ldots, M.
\end{eqnarray}
So the resulting $K_{\underline{t}}$ behaves like
\begin{eqnarray}
K_{\underline{t}} &=& \sum_{i = 1}^M \frac{(t_{i - 1} - t_i)^2}{\tilde{Q}(t_{i - 1}) - \tilde{Q}(t_i)} \rightarrow \int_0^1 \frac{1}{\tilde{Q}^\prime(t)} dt\nonumber\\
&=& 2\sqrt{\pi} \int_0^1 \sqrt{-\log t} dt = 2\sqrt{\pi} \int_{-\infty}^\infty y^2 e^{-y^2}dy = \pi.
\end{eqnarray}
Therefore, as the quantization goes fine asymptotically, the effective SNR as given by (\ref{eqn:eff-snr-mbit}) approaches the actual SNR, and thus the performance loss due to quantization eventually diminishes.

{\it Example:} 4-level quantization, $M = 2$

In this case, there is only one variable, $t = t_1$, to optimize. The maximization problem becomes
\begin{eqnarray}
\max_{t \in (0, 1)} \frac{(1 - t)^2}{1/2 - \tilde{Q}(t)} + \frac{t^2}{\tilde{Q}(t)}.
\end{eqnarray}
A numerical computation immediately gives $\max_{t \in (0, 1)} K_t = 2.7775$, and interestingly, the maximizing $t = 0.618$ is the golden ratio.

{\it Example:} Uniform quantization

In practical systems, uniform quantization is common, in which the thresholds satisfy $\alpha_i = i \sqrt{2(\mathcal{E}_s + \sigma^2)\alpha}$ for $i = 0, 1, \ldots, M - 1$, and $\alpha_M = \infty$, where $\alpha > 0$ is a parameter for optimization. These thresholds lead to
\begin{eqnarray}
K_{\underline{t}} = \sum_{i = 1}^{M - 1} \frac{\left[e^{-(i - 1)^2 \alpha} - e^{-i^2 \alpha}\right]^2}{Q(\sqrt{2\alpha} (i - 1)) - Q(\sqrt{2\alpha} i)} + \frac{e^{-2(M - 1)^2 \alpha}}{Q(\sqrt{2\alpha}(M - 1))},
\end{eqnarray}
which can be further maximized over $\alpha > 0$.

In Table \ref{tab:uniform}, we list the numerical results for optimizing $K_{\underline{t}}$ over $\alpha$, up until $M = 8$.
\begin{table}[ht]
\vspace{0.1in}
\begin{tabular}{|l|l|l|l|l|l|l|l|}
  \hline
  $M$ & 2 & 3 & 4 & 5 & 6 & 7 & 8 \\
  \hline
  $\max_\alpha K_{\underline{t}}$ & 2.7725 & 2.9569 & 3.0291 & 3.0651 & 3.0858 & 3.0989 & 3.1077 \\
  \hline
  optimal $\alpha$ & 0.481 & 0.253 & 0.159 & 0.111 & 0.082 & 0.064 & 0.051 \\
  \hline
\end{tabular}
\vspace{0.1in}
\caption{Table of performance for optimized uniform $2M$-level symmetric output quantization.}
\label{tab:uniform}
\end{table}

{\it Example:} $t$-uniform quantization

An alternative quantizer design is to let the values of $\underline{t}$ be uniformly placed within $[0, 1]$, {\it i.e.}, $t_i = (M - i)/M$, for $i = 0, 1, \ldots, M$. This quantization leads to
\begin{eqnarray}
K_{\underline{t}} = \frac{1}{M^2} \sum_{i = 1}^M \frac{1}{\tilde{Q}(1 - (i - 1)/M) - \tilde{Q}(1 - i/M)}.
\end{eqnarray}

In Table \ref{tab:t-uniform}, we list the numerical results of $K_{\underline{t}}$ for $t$-uniform quantizers, up until $M = 8$. We notice that the $t$-uniform quantization is consistently inferior to the optimized uniform quantization.
\begin{table}[ht]
\vspace{0.1in}
\begin{tabular}{|l|l|l|l|l|l|l|l|}
  \hline
  $M$ & 2 & 3 & 4 & 5 & 6 & 7 & 8 \\
  \hline
  $K_{\underline{t}}$ & 2.7488 & 2.9267 & 3.0011 & 3.0404 & 3.0642 & 3.0798 & 3.0908 \\
  \hline
\end{tabular}
\vspace{0.1in}
\caption{Table of performance for $t$-uniform $2M$-level symmetric output quantization.}
\label{tab:t-uniform}
\end{table}

{\it Example:} Optimal quantization

We can also develop program to numerically solve the optimization problem (\ref{eqn:maximization-Kt}). In Table \ref{tab:optimal}, we list the results, up until $M = 8$. We also list the value of the optimal $t_1$, from which we can recursively compute the whole optimal $\underline{t}$ vector, through enforcing the partial derivatives $\partial{K_{\underline{t}}}/{\partial t_i}$ to vanish for $i = 2, \ldots, M-1$ progressively.
\begin{table}[ht]
\vspace{0.1in}
\begin{tabular}{|l|l|l|l|l|l|l|l|}
  \hline
  $M$ & 2 & 3 & 4 & 5 & 6 & 7 & 8 \\
  \hline
  $\max_{\underline{t}} K_{\underline{t}}$ & 2.7725 & 2.9595 & 3.0330 & 3.0695 & 3.0902 & 3.1032 & 3.1117 \\
  \hline
  optimal $t_1$ & 0.618 & 0.805 & 0.880 & 0.922 & 0.943 & 0.958 & 0.967 \\
  \hline
\end{tabular}
\vspace{0.1in}
\caption{Table of performance for optimal $2M$-level symmetric output quantization.}
\label{tab:optimal}
\end{table}

\begin{figure}[ht]
\centering
\includegraphics[width=3.5in]{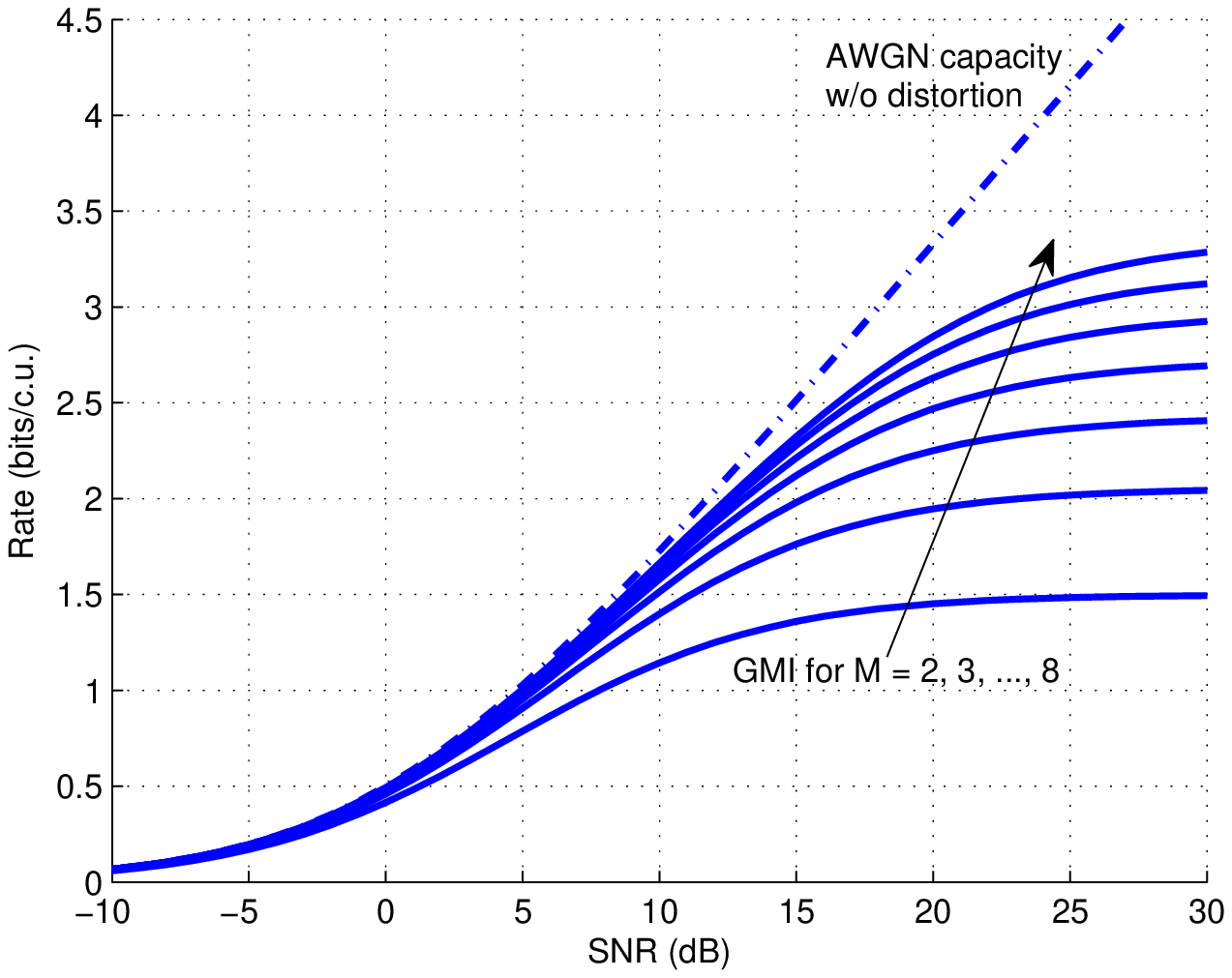}
\caption{The GMI achieved by optimal $2M$-level quantizers, for $M = 2, 3, \ldots, 8$.}
\label{fig:opt-adc-nyquist}
\end{figure}

From the numerical results in the above examples, we observe that the GMI may be fairly close to the channel capacity at low SNR. For example, with the optimal octal quantizer ($M = 4$), the low-SNR GMI scales with SNR like $0.4827 \cdot \mathrm{SNR}$ bits/c.u., which is better than the known lower bound $0.475 \cdot \mathrm{SNR}$ bits/c.u. in the literature \cite{viterbi79:book}. In Figure \ref{fig:opt-adc-nyquist} we plot the GMI $I_\mathrm{GMI}$ achieved by the optimal quantizers, for $M = 2, 3, \ldots, 8$. For comparison we also plot in dash-dot curve the capacity $(1/2)\log_2(1 + \mathrm{SNR})$ of undistorted channels. We can roughly assess that, with $M = 4$ ({\it i.e.}, 3-bit quantization), the performance gap between the GMI and the undistorted channel capacity is mild up until $\mathrm{SNR} \approx 10$ dB; and with $M = 8$ ({\it i.e.}, 4-bit quantization), the performance gap is mild up until $\mathrm{SNR} \approx 15$ dB. Compared with the numerically evaluated capacity for 2/3-bit quantization in \cite{singh09:com}, we see that using the Gaussian codebook ensemble and the nearest-neighbor decoding rule induce a 15-25\% rate loss at high SNR. Comparing Tables \ref{tab:uniform} and \ref{tab:optimal}, we further notice that the performance loss due to using uniform quantization is essentially negligible.

\subsubsection*{Remark on Possible Connection with Capacity per Unit Cost}

For a given $2M$-level symmetric quantizer, we can evaluate the channel capacity per unit cost (symbol energy in our context) by optimizing a single nonzero input symbol, $x$ (see \cite{verdu90:it}). Without loss of generality, we let $x > 0$ and the noise variance $\sigma^2$ be unity. Then the capacity per unit cost can be evaluated as
\begin{eqnarray}
\label{eqn:cpuc}
\sup_{x > 0} \frac{1}{x^2} \sum_{i = 1}^M \left[(Q(\alpha_{i - 1} - x) - Q(\alpha_i - x))\log\frac{Q(\alpha_{i - 1} - x) - Q(\alpha_i - x)}{Q(\alpha_{i - 1}) - Q(\alpha_i)} \right.\nonumber\\
\left.+ (Q(\alpha_{i - 1} + x) - Q(\alpha_i + x))\log\frac{Q(\alpha_{i - 1} + x) - Q(\alpha_i + x)}{Q(\alpha_{i - 1}) - Q(\alpha_i)}\right].
\end{eqnarray}
With some manipulations, we find that $K_{\underline{t}}/(2\pi)$ is exactly the limit value of the term in (\ref{eqn:cpuc}) as $x \rightarrow 0$.\footnote{This is also half of the Fisher information for estimating $\rvx = 0$ from the quantized channel output $\rvw$ \cite{verdu90:it}.} Therefore, only if the capacity per unit cost (\ref{eqn:cpuc}) is achieved by $x \rightarrow 0$, the GMI coincides with the channel capacity in the low-SNR limit. Unfortunately, as revealed by our numerical experiments, this condition does not generally hold for all possible symmetric quantizers.

\section{Super-Nyquist Output Sampling}
\label{sec:super-nyquist}

In this section, we examine the scenario where we sample the channel output at a rate higher than the Nyquist rate, and investigate the benefit of increased  sampling rates in terms of the GMI.

We start with a continuous-time baseband model in which the transmitted signal is
\begin{eqnarray}
x(t) = \frac{1}{\sqrt{2W}} \sum_{k = 1}^n \rvx_k g\left(t - \frac{k}{2W}\right),
\end{eqnarray}
where $g(\cdot)$ is a pulse function with unit energy and is band limited within $W$ Hz. In analysis, a commonly used pulse function is the sinc function $g(t) = \sqrt{2W} \mathrm{sinc}(2Wt)$ with $\mathrm{sinc}(t) = \sin(\pi t)/(\pi t)$, which vanishes at the Nyquist sampling time instants $t = \{k/(2W)\}_{k = -\infty}^\infty$. The channel input is a sequence of independent $\mathcal{N}(0, \mathcal{E}_s)$ random variables $\{\rvx_k\}_{k = 1}^n$. With additive white Gaussian noise $z(t)$, the received signal is
\begin{eqnarray}
y(t) = x(t) + z(t).
\end{eqnarray}
We assume that $z(t)$ is band-limited within $W$ Hz, with in-band two-sided power spectral density $\sigma^2/2$. So the autocorrelation function of $z(t)$ is $K_z(\tau) = \frac{\sigma^2}{2} \mathrm{sinc}(2Wt)$.

We consider a uniform sampler, which samples the channel output $y(t)$ at $L$ times the Nyquist rate. For the $k$-th input symbol, the sampling time instants thus are
\begin{eqnarray}
\label{eqn:sampling-time}
t = \left\{\frac{k}{2W} + \frac{l}{2WL} - \tau_L\right\}_{l = 0}^{2(L - 1)}.
\end{eqnarray}
Here, $\tau_L$ is a constant offset to ensure that the sampling times are symmetric with respect to the center of the $k$-th input symbol pulse; for example, $\tau_1 = 0$ (Nyquist sampling), $\tau_2 = 1/(4W)$, $\tau_3 = 1/(3W)$, $\tau_4 = 3/(8W)$... Generally, $\tau_L = \frac{L - 1}{L} \frac{1}{2W}$. Thus we can rewrite (\ref{eqn:sampling-time}) as
\begin{eqnarray}
t = \frac{1}{2W}\left\{k + \frac{l}{L}\right\}_{l = -L+1}^{L-1}.
\end{eqnarray}
Denote the output samples by $\{\rvy_{k, l}\}$ with $\rvy_{k, l} = y(t_{k, l})$ where $t_{k, l} = \frac{1}{2W}(k + l/L)$. The samples pass through a nonlinear distortion device, so that the observed samples are $\rvw_{k, l} = f(\rvy_{k, l})$.

Let us generalize the nearest-neighbor decoding rule in Section \ref{sec:basic-model-real} as follows. For all possible messages, the decoder computes the distance metrics,
\begin{eqnarray}
D(m) = \frac{1}{n}\sum_{k = 1}^n \sum_{l = -L+1}^{L-1} \xi_l [w_{k, l} - a_l x_k(m)]^2, \quad m \in \mathcal{M},
\end{eqnarray}
where $\{\xi_l\}_{l = -L+1}^{L-1}$ and $\{a_l\}_{l = -L+1}^{L-1}$ are weighting coefficients, and decides the received message as $\hat{m} = \mathrm{arg}\min_{m \in \mathcal{M}} D(m)$. We then note that
\begin{eqnarray*}
&&\sum_{l = -L+1}^{L-1} \xi_l [w_{k, l} - a_l x_k(m)]^2 = \sum_{l = -L+1}^{L-1} \xi_l w_{k, l}^2 - 2 x_k(m) \sum_{l = -L+1}^{L-1} \xi_l a_l w_{k, l} + x_k^2(m) \sum_{l = -L+1}^{L-1} \xi_l a_l^2\nonumber\\
&=& \left(\sum_{l = -L+1}^{L-1} \xi_l a_l^2\right)\cdot \left[x_k(m) - \frac{\sum_{l = -L+1}^{L-1} \xi_l a_l w_{k, l}}{\sum_{l = -L+1}^{L-1} \xi_l a_l^2}\right]^2 + \left[\sum_{l = -L+1}^{L-1} \xi_l w_{k, l}^2 - \frac{\left(\sum_{l = -L+1}^{L-1} \xi_l a_l w_{k, l}\right)^2}{\sum_{l = -L+1}^{L-1} \xi_l a_l^2}\right].
\end{eqnarray*}
Therefore, without loss of generality, we may consider the simplified nearest-neighbor decoding distance metric
\begin{eqnarray}
D(m) = \frac{1}{n} \sum_{k = 1}^n \left[\sum_{l = -L+1}^{L-1} \beta_l w_{k, l} - x_k(m)\right]^2,
\end{eqnarray}
for which the tunable weighting coefficients are $\underline{\beta} = \{\beta_l\}_{l = -L+1}^{L-1}$.

Following the same procedure as that in Section \ref{sec:basic-model-real} for the Nyquist-sampled channel model, we first examine the limit of $D(1)$ assuming that the message $m = 1$ is sent. Since the channel input symbols $\rvx_\cdot$ are i.i.d. and the noise process is wide-sense stationary, the observed samples $\rvw_{k, l}$ constitute an ergodic process.\footnote{We note that the transmission of a codeword, $\{\rvx_k\}_{k = 1}^n$, is finite-length. In order to meet the ergodicity condition, we may slightly modify the model by appending $\{\rvx_k\}_{k = -\infty}^0$ and $\{\rvx_k\}_{k = n + 1}^\infty$, which consist of i.i.d. $\mathcal{N}(0, \mathcal{E}_s)$ random variables as additional interference, to the transmitted codeword.} Consequently, we have
\begin{eqnarray}
\lim_{n \rightarrow \infty} D(1) = \mathbf{E} \left[\sum_{l = -L+1}^{L-1} \beta_l \rvw_{0, l} - \rvx_0\right]^2 \quad \mbox{a.s.}
\end{eqnarray}
On the other hand, for any $m \neq 1$, we have
\begin{eqnarray}
\frac{1}{n} \Lambda_n(n\theta) 
&=& \frac{1}{n} \log \mathbf{E} \left\{\left.e^{\theta \sum_{k = 1}^n \left[\sum_{l = -L+1}^{L - 1} \beta_l \rvw_{k, l} - \rvx_k(m)\right]^2}\right| \rvw_{k, l}, k = 1, \ldots, n, l = -L+1,
\ldots, L - 1\right\}\nonumber\\
&=& \frac{\theta}{1 - 2\theta \mathcal{E}_s} \frac{1}{n} \sum_{k = 1}^n \left[\sum_{l = -L+1}^{L - 1} \beta_l \rvw_{k, l}\right]^2 - \frac{1}{2} \log(1 - 2\theta \mathcal{E}_s)\nonumber\\
&\rightarrow& \frac{\theta}{1 - 2\theta \mathcal{E}_s} \mathbf{E} \left[\sum_{l = -L+1}^{L - 1} \beta_l \rvw_{0, l}\right]^2 - \frac{1}{2} \log(1 - 2\theta \mathcal{E}_s) \quad \mbox{a.s.}
\end{eqnarray}
In both limits above, $\{\rvw_{0, l}\}_{l = -L+1}^{L - 1}$ are induced by an infinite sequence of inputs, $\{\rvx_k\}_{k = -\infty}^\infty$.

So the GMI is
\begin{eqnarray}
\label{eqn:i-gmi-super-nyquist}
I_\mathrm{GMI} = \sup_{\underline{\beta}, \theta < 0} \left\{\theta \mathbf{E} \left[\sum_{l = -L+1}^{L - 1} \beta_l \rvw_{0, l} - \rvx_0\right]^2 - \frac{\theta}{1 - 2\theta \mathcal{E}_s} \mathbf{E} \left[\sum_{l = -L+1}^{L - 1} \beta_l \rvw_{0, l}\right]^2 + \frac{1}{2} \log(1 - 2\theta \mathcal{E}_s) \right\},
\end{eqnarray}
and we have the following result, whose derivation is given in Supplementary Material \ref{appendix:proof-gmi-super-nyquist}.
\begin{prop}
\label{prop:gmi-super-nyquist}
The GMI with super-Nyquist output sampling is
\begin{eqnarray}
\label{eqn:gmi-super-nyquist}
I_\mathrm{GMI} = \frac{1}{2}\log\left(1 + \frac{\Delta}{1 - \Delta}\right),
\end{eqnarray}
where $\Delta = \left(\underline{b}^T \mathbf{\Omega}^{-1} \underline{b}\right)/{\mathcal{E}_s}$, $\mathbf{\Omega}$ is a $(2L-1) \times (2L-1)$ matrix with its $(u, l)$-element being $\mathbf{E}[\rvw_{0, u} \rvw_{0, l}]$, and $\underline{b}$ is a $(2L-1)$-dimensional vector with its $l$-element being $\mathbf{E}[\rvx_0 \rvw_{0, l}]$, $u, l = -L+1, \ldots, L-1$. To achieve the GMI in (\ref{eqn:gmi-super-nyquist}), the optimal weighting coefficients are
\begin{eqnarray}
\underline{\beta} = \frac{\mathcal{E}_s}{\underline{b}^T \mathbf{\Omega}^{-1} \underline{b}} \mathbf{\Omega}^{-1} \underline{b}.
\end{eqnarray}
\end{prop}

We notice that the GMI in (\ref{eqn:gmi-super-nyquist}) is a natural extension of that in Proposition \ref{prop:gmi-real} for the Nyquist-sampling case, and we can also define the effective SNR by $\mathrm{SNR}_e = \Delta/(1 - \Delta)$.

\subsection{Binary Symmetric Quantization: Sinc Pulse Function}

We examine binary symmetric quantization in which $w = \mathrm{sgn}(y)$. For this purpose, we need to evaluate $\mathbf{\Omega}$ and $\underline{b}$. For each $l$,
\begin{eqnarray}
\rvy_{0, l} = \frac{1}{\sqrt{2W}} \sum_{k = -\infty}^\infty \rvx_k g\left(\frac{l}{2WL} - \frac{k}{2W}\right) + \rvz\left(\frac{l}{2WL}\right).
\end{eqnarray}

Utilizing (\ref{eqn:x-sgny}) and noting that $\{\rvx_k\}$ are i.i.d., we have
\begin{eqnarray}
\label{eqn:b-vector}
b_l &=& \mathbf{E}[\rvx_0 \mathrm{sgn}(\rvy_{0, l})]\nonumber\\
&=& \frac{\mathcal{E}_s g(l/(2WL))}{\sqrt{\pi \left[(\mathcal{E}_s/2)\sum_{k = -\infty}^\infty g^2(l/(2WL) - k/(2W)) + \sigma^2 W/2\right]}},
\end{eqnarray}
for $l = -L+1, \ldots, L - 1$.

The undistorted received signal samples, $\rvy_{0, u}$ and $\rvy_{0, l}$, are jointly zero-mean Gaussian. We can further evaluate their correlation as
\begin{eqnarray}
\label{eqn:r-matrix}
&&r_{u, l} = \frac{\mathbf{E} [\rvy_{0, u} \rvy_{0, l}]}{\sqrt{\mathrm{var}[\rvy_{0, u}]}\cdot \sqrt{\mathrm{var}[\rvy_{0, l}]}}\nonumber\\
&=& \frac{\frac{\mathcal{E}_s}{2W}\sum_{k = -\infty}^\infty g\left(l/(2WL) - k/(2W)\right) g\left(u/(2WL) - k/(2W)\right) + \frac{\sigma^2}{2} \mathrm{sinc}\left((l - u)/L\right)}{\sqrt{\frac{\mathcal{E}_s}{2W} \sum_{k = -\infty}^\infty g^2(l/(2WL) - k/(2W)) + \frac{\sigma^2}{2}} \sqrt{\frac{\mathcal{E}_s}{2W} \sum_{k = -\infty}^\infty g^2(u/(2WL) - k/(2W)) + \frac{\sigma^2}{2}}}.\nonumber
\end{eqnarray}
Consequently, the correlation between the hard-limited samples is \cite{vanvleck66:pieee}
\begin{eqnarray}
\Omega_{u, l} = \mathbf{E}[\rvw_{0, u} \rvw_{0, l}] = \frac{2}{\pi} \arcsin r_{u, l}.
\end{eqnarray}

Now in this subsection we focus on the sinc pulse function, $g(t) = \sqrt{2W} \mathrm{sinc}(2Wt)$. For this $g(\cdot)$, through (\ref{eqn:b-vector}) and (\ref{eqn:r-matrix}) we have
\begin{eqnarray}
b_l &=& \frac{2\mathcal{E}_s}{\sqrt{\pi \sigma^2}} \frac{\mathrm{sinc}(l/L)}{\sqrt{(2\mathcal{E}_s/\sigma^2) \Xi(l, l) + 1}},\\
r_{u, l} &=& \frac{(2 \mathcal{E}_s/\sigma^2) \Xi(l, u) + \mathrm{sinc}\left(\frac{l - u}{L}\right)}{\sqrt{(2 \mathcal{E}_s/\sigma^2)\Xi(l, l) + 1} \sqrt{(2 \mathcal{E}_s/\sigma^2) \Xi(u, u) + 1}},
\end{eqnarray}
where $\Xi(l, u) = \sum_{k = -\infty}^\infty \mathrm{sinc}\left(l/L - k\right) \mathrm{sinc}\left(u/L - k\right)$, which can be further evaluated as $\Xi(l, u) = \mathrm{sinc}((l - u)/L)$, for all $l, u = -L+1, \ldots, L - 1$. So
\begin{eqnarray}
b_l = \sqrt{\frac{2\mathcal{E}_s}{\pi}} \sqrt{\frac{2\mathcal{E}_s/\sigma^2}{2\mathcal{E}_s/\sigma^2 + 1}} \mathrm{sinc}\left(l/L\right), \quad\mbox{and}\; r_{u, l} = \mathrm{sinc}\left(\frac{l - u}{L}\right).
\end{eqnarray}

When $L = 1$, {\it i.e.}, Nyquist sampling, we can easily verify that $\Delta = \frac{2}{\pi} \frac{\mathcal{E}_s}{\mathcal{E}_s + \sigma^2/2}$, thus revisiting the result in Section \ref{sec:1-bit}.

From the above, we can find the following behavior of the GMI, in which we denote $\mathrm{SNR} = \frac{\mathcal{E}_s}{\sigma^2/2}$, $\underline{b}_0 = \left[\mathrm{sinc}\left(l/L\right)\right]_{l = -L+1, \ldots, L - 1}$, and $\mathbf{\Omega}_0 = \left[\arcsin \mathrm{sinc}\left((l - u)/L\right)\right]_{l, u = -L+1, \ldots, L - 1}$.
\begin{eqnarray}
\Delta = \frac{\mathrm{SNR}}{\mathrm{SNR} + 1} \underline{b}_0^T \mathbf{\Omega}_0^{-1} \underline{b}_0,\quad\mbox{and}\; \mathrm{SNR}_e = \frac{\underline{b}_0^T \mathbf{\Omega}_0^{-1} \underline{b}_0 \cdot \mathrm{SNR}}{(1 - \underline{b}_0^T \mathbf{\Omega}_0^{-1} \underline{b}_0)\cdot \mathrm{SNR} + 1}.
\end{eqnarray}
\begin{itemize}
\item High-SNR regime: As $\mathrm{SNR} \rightarrow \infty$,
\begin{eqnarray}
I_\mathrm{GMI} = \frac{1}{2} \log \left(\frac{1}{1 - \underline{b}_0^T \mathbf{\Omega}_0^{-1} \underline{b}_0}\right) + o(1).
\end{eqnarray}
\item Low-SNR regime: As $\mathrm{SNR} \rightarrow 0$,
\begin{eqnarray}
I_\mathrm{GMI} = \frac{\underline{b}_0^T \mathbf{\Omega}_0^{-1} \underline{b}_0}{2} \mathrm{SNR} + o(\mathrm{SNR}).
\end{eqnarray}
\end{itemize}

In Table \ref{tab:super-nyquist}, we present the numerical results for the asymptotic behavior of the GMI, for different values of $L$.
\begin{table}[ht]
\vspace{0.1in}
\begin{tabular}{|l|l|l|l|l|l|l|l|}
  \hline
  $L$ & 1 & 2 & 4 & 8 & 16 & 32 & $\infty$ \\
  \hline
  $\underline{b}_0^T \mathbf{\Omega}_0^{-1} \underline{b}_0$ & $2/\pi$ & 0.7173 & 0.7591 & 0.7734 & 0.7783 & 0.7801 & 0.7815 \\
  \hline
  $\lim_{\mathrm{SNR} \rightarrow \infty} I_\mathrm{GMI}$ (bits/c.u.) & 0.7302 & 0.9114 & 1.0268 & 1.0710 & 1.0867 & 1.0926 & 1.0970 \\
  \hline
  $\lim_{\mathrm{SNR} \rightarrow 0} I_\mathrm{GMI}/\mathrm{SNR}$ & 0.3183 & 0.3587 & 0.3796 & 0.3867 & 0.3892 & 0.3901 & 0.3907 \\
  \hline
\end{tabular}
\vspace{0.1in}
\caption{Table of performance for super-Nyquist output sampling with binary symmetric quantization and sinc pulse function.}
\label{tab:super-nyquist}
\end{table}
From the numerical results, we see that super-Nyquist sampling yields noticeable benefit for the GMI. In the low-SNR regime, sampling at twice the Nyquist rate attains $\lim_{\mathrm{SNR} \rightarrow 0} I_\mathrm{GMI}/\mathrm{SNR} = 0.3587$, which is slightly smaller than the lower bound $0.3732$ which has been obtained in \cite{koch10:arxiv}. In the high-SNR regime, we further observe that for $L \geq 4$ the GMI exceeds $1$ bit/c.u.! Intuitively, this is due to the fact that the diversity yielded by super-Nyquist sampling is capable of exploiting the abundant information carried by the Gaussian codebook ensemble.

\begin{figure}[ht]
\centering
\includegraphics[width=3.5in]{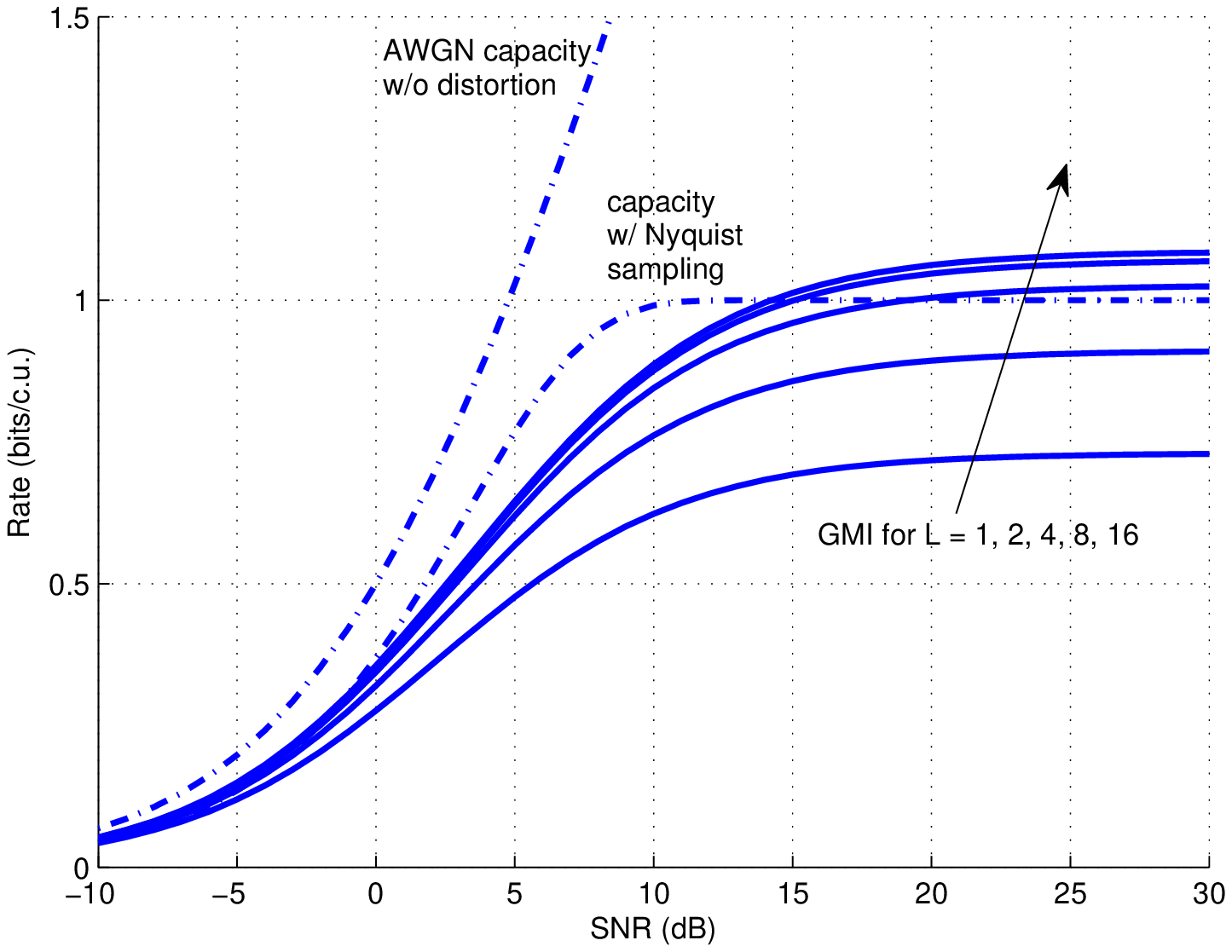}
\caption{The GMI achieved by super-Nyquist sampling with binary symmetric quantization and sinc pulse function, for $L = 1, 2, 4, 8, 16$.}
\label{fig:super-nyquist}
\end{figure}

To further consolidate our above analysis, in Figure \ref{fig:super-nyquist} we plot the GMI achieved for different values of $L$. We can clearly observe the rate gain by increasing the sampling rate. For comparison, we also plot the AWGN capacity without distortion and the capacity under binary symmetric quantization and with Nyquist sampling \cite{singh09:com}. We notice that, as $L$ increases, on one hand, the performance gap between the GMI and the capacity tends to diminish for SNR smaller than 0 dB; on the other hand, the GMI even outperforms the capacity at high SNR.

\subsection{Binary Symmetric Quantization: Pulse Function Optimization at Low SNR}

We have already seen in the previous subsection that super-Nyquist sampling yields noticeable benefit. In this subsection, we illustrate that we can even realize additional benefit through optimizing the pulse function $g(\cdot)$.

With sampling factor $L$, we restrict the pulse function to take the following form
\begin{eqnarray}
g(t) = \sum_{v = -L+1}^{L-1} \gamma_v \sqrt{2W} \mathrm{sinc}(2Wt - v/L);
\end{eqnarray}
that is, a superposition of $2L - 1$ (time-shifted) sinc pulses. The weighting parameters $\{\gamma_v\}_{v = -L+1}^{L-1}$ are such that the energy of $g(t)$ is unity, {\it i.e.},
\begin{eqnarray}
\int_{-\infty}^\infty g^2(t) dt = \sum_{v = -L+1}^{L-1} \sum_{v^\prime = -L+1}^{L-1} \gamma_v \gamma_{v^\prime} \mathrm{sinc}\left(\frac{v - v^\prime}{L}\right) = 1,
\end{eqnarray}
which may be rewritten in matrix form as $\underline{\gamma}^T \mathbf{\Theta} \underline{\gamma} = 1$, where $\mathbf{\Theta} = \left[\mathrm{sinc}\left((l - u)/L\right)\right]_{l, u = -L+1, \ldots, L - 1}$. If we let $\gamma_0 = 1$ and $\gamma_{v \neq 0} = 0$, we obtain the sinc pulse function.

Through the general formulas (\ref{eqn:b-vector}) and (\ref{eqn:r-matrix}), we have, after some algebraic manipulation,
\begin{eqnarray}
b_l &=& \sqrt{\frac{2\mathcal{E}_s}{\pi}} \sqrt{\frac{2\mathcal{E}_s/\sigma^2}{2\mathcal{E}_s/\sigma^2 + 1}} \sum_{v = -L+1}^{L-1} \gamma_v \mathrm{sinc}\left(\frac{l - v}{L}\right),\\
r_{u, l} &=& \frac{(2\mathcal{E}_s/\sigma^2) \sum_{a = -L+1}^{L-1}\sum_{b = -L+1}^{L-1} \gamma_a \gamma_b \mathrm{sinc}\left(\frac{l - u - a + b}{L}\right) + \mathrm{sinc}\left(\frac{l - u}{L}\right)}{2\mathcal{E}_s/\sigma^2 + 1}.
\end{eqnarray}

To illustrate the benefit of optimizing the pulse function, we focus on the low-SNR regime, where $\mathrm{SNR} = \frac{\mathcal{E}_s}{\sigma^2/2}$ approaches toward zero. We thus have
\begin{eqnarray}
\sqrt{\frac{\pi}{2\mathcal{E}_s}} \frac{b_l}{\sqrt{\mathrm{SNR}}} \rightarrow \sum_{v = -L+1}^{L-1} \gamma_v \mathrm{sinc}\left(\frac{l - v}{L}\right), \quad\mbox{and}\;\; r_{u, l} &\rightarrow& \mathrm{sinc}\left(\frac{l - u}{L}\right).
\end{eqnarray}
Subsequently, the value of $\Delta$ and $\mathrm{SNR}_e$ in Proposition \ref{prop:gmi-super-nyquist} behaves like
\begin{eqnarray}
\lim_{\mathrm{SNR} \rightarrow 0}\frac{\mathrm{SNR}_e}{\mathrm{SNR}} = \lim_{\mathrm{SNR} \rightarrow 0} \frac{\Delta}{\mathrm{SNR}} = \underline{\gamma}^T \mathbf{\Theta} \mathbf{\Omega}_0^{-1} \mathbf{\Theta} \underline{\gamma},
\end{eqnarray}
where $\mathbf{\Theta} = \left[\mathrm{sinc}\left((l - u)/L\right)\right]_{l, u = -L+1, \ldots, L - 1}$ and $\mathbf{\Omega}_0 = \left[\arcsin \mathrm{sinc}\left((l - u)/L\right)\right]_{l, u = -L+1, \ldots, L - 1}$ have been defined previously. Keeping in mind the unit-energy constraint on $g(t)$, the following optimization problem is immediate,
\begin{eqnarray}
\max_{\underline{\gamma}} \underline{\gamma}^T \mathbf{\Theta} \mathbf{\Omega}_0^{-1} \mathbf{\Theta} \underline{\gamma}, \quad \mbox{s.t.}\;\; \underline{\gamma}^T \mathbf{\Theta} \underline{\gamma} = 1.
\end{eqnarray}
By noting that $\mathbf{\Theta}$ is a positive-definite matrix, we can introduce the transform $\tilde{\underline{\gamma}} = \mathbf{\Theta}^{1/2} \underline{\gamma}$, and rewrite the optimization problem as
\begin{eqnarray}
\max_{\underline{\gamma}} \frac{\tilde{\underline{\gamma}}^T \mathbf{\Theta}^{1/2} \mathbf{\Omega}_0^{-1} \mathbf{\Theta}^{1/2} \tilde{\underline{\gamma}}}{\tilde{\underline{\gamma}}^T \tilde{\underline{\gamma}}},
\end{eqnarray}
for which the maximum value is the largest eigenvalue of $\mathbf{\Theta}^{1/2} \mathbf{\Omega}_0^{-1} \mathbf{\Theta}^{1/2}$, and the optimal $\tilde{\underline{\gamma}}$ is the unit-norm eigenvector corresponding to the largest eigenvalue.

In Table \ref{tab:super-nyquist-opt-pulse}, we present the numerical results for the low-SNR asymptotic behavior of the GMI, with the optimal choice of $\underline{\gamma}$, for different values of $L$.
\begin{table}[ht]
\vspace{0.1in}
\begin{tabular}{|l|l|l|l|l|l|l|}
  \hline
  $L$ & 2 & 4 & 8 & 16 & 32 & $\infty$ \\
  \hline
  $\lim_{\mathrm{SNR} \rightarrow 0} I_\mathrm{GMI}/\mathrm{SNR}$ & 0.3731 & 0.3923 & 0.3971 & 0.3984 & 0.3987 & 0.3988 \\
  \hline
\end{tabular}
\vspace{0.1in}
\caption{Table of performance for super-Nyquist output sampling with binary symmetric quantization and optimized pulse function.}
\label{tab:super-nyquist-opt-pulse}
\end{table}
Compared with Table \ref{tab:super-nyquist}, we notice that optimizing the pulse function leads to a noticeable additional improvement on the GMI. In particular, for $L = 2$ our approach yields $\lim_{\mathrm{SNR} \rightarrow 0} I_\mathrm{GMI}/\mathrm{SNR} = 0.3731$, which almost coincides with the result in \cite{koch10:arxiv}, $0.3732$.\footnote{Since both our result and that in \cite{koch10:arxiv} are analytical, we have compared their values in fine precision and found that they are indeed different.}

\section{Conclusions}
\label{sec:conclusions}

With the surging quest for energy-efficient communication solutions, transceivers with deliberately engineered distortions have attracted much attention in system design. These distortions, such as transmit-side clipping and low-precision receive-side quantization, may significantly alleviate power consumption and hardware cost. It is thus imperative for communication engineers to develop a systematic understanding of the impact of these distortions, so as to assess the resulting system performance, and to guide the design of distortion mechanisms. In this paper, we make an initial attempt at this goal, developing a general analytical framework for evaluating the achievable information rates using the measure of GMI, and illustrating the application of this framework by examining several representative transceiver distortion models. We hope that both the framework and the applications presented in this paper will be useful for deepening our understanding in this area.

Admittedly, the approach taken in this paper, namely evaluating the GMI for Gaussian codebook ensemble and nearest-neighbor decoding, is inherently suboptimal for general transceiver distortion models. Nevertheless, as illustrated throughout this paper, the general analytical framework built upon such an approach is convenient for performance evaluation and instrumental for system design. In many practically important scenarios, for example the low/moderate-SNR regime, this approach leads to near-optimal performance. Furthermore, as suggested by our analysis of super-Nyquist sampling, we can substantially alleviate the performance loss by sampling the channel output at rates higher than the Nyquist rate.

A number of interesting problems remain unsolved within the scope of this paper. These include, among others: answering whether the GMI coincides with the channel capacity for multi-bit output quantization in the low-SNR limit; identifying more effective ways of processing the samples in super-Nyquist sampled channels; characterizing the ultimate performance limit of super-Nyquist sampling. Beyond the scope of this paper, one can readily see a whole agenda of research on communication with nonlinear transceiver distortion, including timing recovery, channel estimation, equalization, transmission under multipath fading, and multiantenna/multiuser aspects.

\bibliographystyle{ieee}
\bibliography{./nonlinear}

\newpage
\section*{Supplementary Material}

\subsection{Derivation of the GMI in Proposition \ref{prop:gmi-real}}
\label{appendix:proof-gmi-real}

We proceed starting from (\ref{eqn:Lambda_n}) as follows. For any $m \neq 1$,
\begin{eqnarray}
&&\mathbf{E}\left\{\left.e^{n\theta D(m)}\right| \rvw_k, k = 1, \ldots, n \right\} = \mathbf{E}\left\{\left.e^{\theta \sum_{k = 1}^n [\rvw_k - a \rvx_k(m)]^2}\right| \rvw_k, k = 1, \ldots, n \right\}\nonumber\\
&=& \prod_{k = 1}^n \mathbf{E}\left\{\left.e^{\theta [\rvw_k - a \rvx_k(m)]^2}\right| \rvw_k\right\} = \prod_{k = 1}^n \frac{1}{\sqrt{1 - 2\theta a^2 \mathcal{E}_s}} \exp\left(\frac{\theta \rvw_k^2}{1 - 2\theta a^2 \mathcal{E}_s}\right)\nonumber\\
&=& (1 - 2\theta a^2 \mathcal{E}_s)^{-n/2} \exp\left(\sum_{k = 1}^n \frac{\theta \rvw_k^2}{1 - 2\theta a^2 \mathcal{E}_s}\right),
\end{eqnarray}
by noting that conditioned upon $\rvw_\cdot$, $(\rvw_\cdot - a \rvx_\cdot)^2$ is a noncentral chi-square random variable. This leads to
\begin{eqnarray}
\Lambda_n(n\theta) = \log \mathbf{E}\left\{\left.e^{n\theta D(m)}\right| \rvw_k, k = 1, \ldots, n \right\} = \frac{\theta}{1 - 2\theta a^2 \mathcal{E}_s} \sum_{k = 1}^n \rvw_k^2 - \frac{n}{2} \log(1 - 2\theta a^2 \mathcal{E}_s).
\end{eqnarray}
Consequently, from the law of large numbers,
\begin{eqnarray}
\Lambda(\theta) = \lim_{n \rightarrow \infty} \frac{1}{n} \Lambda_n(n \theta) = \frac{\theta \mathbf{E} \left[f(\rvx, \rvz)\right]^2}{1 - 2\theta a^2 \mathcal{E}_s} - \frac{1}{2} \log(1 - 2\theta a^2 \mathcal{E}_s) \quad \mbox{a.s.}
\end{eqnarray}
where $\rvx \sim \mathcal{N}(0, \mathcal{E}_s)$ and $\rvz \sim \mathcal{N}(0, \sigma^2)$. So we can evaluate the GMI through
\begin{eqnarray}
I_{\mathrm{GMI}} = \sup_{a \in \mathbb{R}, \theta < 0} \left(\theta \mathbf{E}\left\{\left[f(\rvx, \rvz) - a\rvx\right]^2\right\} - \frac{\theta \mathbf{E} \left[f(\rvx, \rvz)\right]^2}{1 - 2\theta a^2 \mathcal{E}_s} + \frac{1}{2}\log(1 - 2\theta a^2 \mathcal{E}_s) \right).
\end{eqnarray}
Note that in the problem formulation we include the optimization of $I_{\mathrm{GMI}}$ over $a \in \mathbb{R}$.

To solve the optimization problem, we define
\begin{eqnarray}
J(a, \theta) &=& \theta \mathbf{E}\left\{\left[f(\rvx, \rvz) - a\rvx\right]^2\right\} - \frac{\theta \mathbf{E} \left[f(\rvx, \rvz)\right]^2}{1 - 2\theta a^2 \mathcal{E}_s} + \frac{1}{2}\log(1 - 2\theta a^2 \mathcal{E}_s)\nonumber\\
&=& \theta \left\{\mathbf{E}[f(\rvx, \rvz)]^2 + a^2 \mathcal{E}_s - 2a \mathbf{E}\left[ f(\rvx, \rvz) \rvx \right]\right\} - \frac{\theta \mathbf{E} [f(\rvx, \rvz)]^2}{1 - 2\theta a^2 \mathcal{E}_s} + \frac{1}{2}\log(1 - 2\theta a^2 \mathcal{E}_s)\nonumber\\
&=& \theta a^2 \mathcal{E}_s + \frac{1}{2}\log(1 - 2\theta a^2 \mathcal{E}_s) - \frac{2\theta^2 a^2 \mathcal{E}_s}{1 - 2\theta a^2 \mathcal{E}_s} \mathbf{E}[f(\rvx, \rvz)]^2 - 2 \theta a \mathbf{E}\left[ f(\rvx, \rvz) \rvx \right].
\end{eqnarray}
By introducing the new variable $\gamma = -2\theta a^2 \mathcal{E}_s > 0$, we rewrite $J(a, \theta)$ as
\begin{eqnarray}
J(\gamma, \theta) = \frac{1}{2} \log(1 + \gamma) - \frac{\gamma}{2} + \frac{\gamma \theta}{1 + \gamma} \mathbf{E}[f(\rvx, \rvz)]^2 + \sqrt{\frac{-2\gamma \theta}{\mathcal{E}_s}} \mathbf{E}\left[ \left|f(\rvx, \rvz) \rvx \right|\right].
\end{eqnarray}

Letting the partial derivative $\partial J/\partial \theta$ be zero, we find that the optimal value of $\theta < 0$ should be
\begin{eqnarray}
\sqrt{-\theta_\mathrm{opt}} = \frac{(1 + \gamma) \mathbf{E}\left[\left| f(\rvx, \rvz) \rvx \right|\right]}{\mathbf{E}[f(\rvx, \rvz)]^2 \sqrt{2\mathcal{E}_s \gamma}}.
\end{eqnarray}
Substituting $\theta_\mathrm{opt}$ into $J(\gamma, \theta)$ followed by some algebraic manipulation, we obtain
\begin{eqnarray}
J(\gamma, \theta_\mathrm{opt}) = \frac{1}{2}\log(1 + \gamma) - \frac{\gamma}{2} + \frac{(1 + \gamma)\left\{ \mathbf{E}\left[ f(\rvx, \rvz) \rvx \right]\right\}^2}{2\mathcal{E}_s \mathbf{E}[f(\rvx, \rvz)]^2}.
\end{eqnarray}
Let us define
\begin{eqnarray}
\Delta = \frac{\left\{ \mathbf{E}[ f(\rvx, \rvz) \rvx ]\right\}^2}{\mathcal{E}_s \mathbf{E}[f(\rvx, \rvz)]^2},
\end{eqnarray}
and maximize $J(\gamma, \theta_\mathrm{opt}) = \frac{1}{2}\log(1 + \gamma) - \frac{\gamma}{2} + (1 + \gamma) \frac{\Delta}{2}$ over $\gamma > 0$. From Cauchy-Schwartz inequality, we see that $\Delta$ is upper bounded by one. It is then straightforward to show that the optimal value of $\gamma$ is $\gamma_\mathrm{opt} = {\Delta}/({1 - \Delta})$, and hence $J(\gamma_\mathrm{opt}, \theta_\mathrm{opt}) = -\frac{1}{2}\log(1 - \Delta)$.

Therefore, the maximum value $J(\gamma_\mathrm{opt}, \theta_\mathrm{opt})$, {\it i.e.}, the GMI, is
\begin{eqnarray}
I_\mathrm{GMI} = \frac{1}{2}\log \left(1 + \frac{\Delta}{1 - \Delta}\right),
\end{eqnarray}
and the optimal choice of the decoding scaling parameter $a$ is $a_\mathrm{opt} = \mathbf{E}\left[ f(\rvx, \rvz) \rvx \right]/\mathcal{E}_s$.

\subsection{Derivation of the GMI for Antipodal Codebook Ensemble}
\label{appendix:antipodal}

We follow the same line of analysis as that for the Gaussian codebook ensemble. For $m = 1$,
\begin{eqnarray}
\lim_{n \rightarrow \infty} D(1) &=& \mathbf{E}\left\{\left[\rvw - a \rvx\right]^2\right\}\nonumber\\
&=& \mathbf{E} [\rvw^2] + a^2 \mathcal{E}_s - 2a \mathbf{E}[\rvw \rvx] \quad \mbox{a.s.}
\end{eqnarray}
where $\rvw = f(\rvx, \rvz)$ denotes the distorted channel output. On the other hand, for any $m \neq 1$, we find that
\begin{eqnarray}
\frac{1}{n}\Lambda_n(n\theta) &=& \frac{\theta}{n}\sum_{k = 1}^n \rvw_k^2 + \theta a^2 \mathcal{E}_s + \frac{1}{n}\sum_{k = 1}^n \log \cosh (2\theta a \sqrt{\mathcal{E}_s} \rvw_k),\\
\mbox{and}\quad \Lambda(\theta) &=& \lim_{n \rightarrow \infty} \frac{1}{n}\Lambda_n(n\theta) = \theta \mathbf{E}[\rvw^2] + \theta a^2 \mathcal{E}_s + \mathbf{E}\log\cosh (2\theta a \sqrt{\mathcal{E}_s} \rvw), \quad \mbox{a.s.}
\end{eqnarray}
Consequently, we can evaluate the GMI by solving
\begin{eqnarray}
I_\mathrm{GMI} = \sup_{\theta < 0, a \in \mathbb{R}} \left(-2 \theta a \mathbf{E}[\rvx f(\rvx, \rvz)] - \mathbf{E}\log\cosh(2\theta a \sqrt{\mathcal{E}_s} f(\rvx, \rvz))\right).
\end{eqnarray}
By letting $-2 \theta a$ be a single variable $t$, we obtain the problem formulation as given by (\ref{eqn:gmi-antipodal}), and by using the first derivative condition for optimality, we obtain the equation for the optimal value of $t$ as given by (\ref{eqn:gmi-antipodal-t}).

\subsection{General Framework for Complex-Valued Nyquist-Sampled Channels}
\label{appendix:basic-model-complex}

We can extend the general GMI formula (\ref{eqn:gmi-real}) for real-valued channels to complex-valued channels. Let the noise $\rvz_\cdot$ be a sequence of i.i.d. circularly symmetric complex Gaussian random variables ({\it i.e.}, $\rvz_\cdot \sim \mathcal{CN}(0, \sigma^2)$). The memoryless nonlinearity mapping $f(\cdot)$ transforms $(x, z)$ into a complex number $f(x, z)$. Hence the observation is $\rvw_k = f(\rvx_k, \rvz_k)$, for $k = 1, 2, \ldots, n$.

For transmission, we restrict the codebook to be an i.i.d. $\mathcal{CN}(0, \mathcal{E}_s)$ ensemble. The decoder follows a nearest-neighbor rule, which computes for all possible messages, the distance metric,
\begin{eqnarray}
\label{eqn:nndec-complex}
D(m) = \frac{1}{n}\sum_{k = 1}^n \left| w_k - a x_k(m)\right|^2, \quad m \in \mathcal{M},
\end{eqnarray}
and decides the received message as $\hat{m} = \mathrm{arg}\min_{m \in \mathcal{M}} D(m)$.

Analogously to the development for the real-valued channel model in Section \ref{sec:basic-model-real}, we arrive at
\begin{eqnarray}
\label{eqn:gmi-opt-complex}
I_{\mathrm{GMI}} = \sup_{a \in \mathbb{C}, \theta < 0} \left(\theta \mathbf{E}\left\{\left|f(\rvx, \rvz) - a\rvx\right|^2\right\} - \frac{\theta \mathbf{E} |f(\rvx, \rvz)|^2}{1 - \theta |a|^2 \mathcal{E}_s} + \log(1 - \theta |a|^2 \mathcal{E}_s) \right).
\end{eqnarray}
Note that in the problem formulation we include the optimization of $I_{\mathrm{GMI}}$ over $a \in \mathbb{C}$.

Define the expression in the right-hand side of (\ref{eqn:gmi-opt-complex}) as $J(a, \theta)$, which can be rewritten as
\begin{eqnarray}
J(a, \theta) = \theta |a|^2 \mathcal{E}_s + \log(1 - \theta |a|^2 \mathcal{E}_s) - \frac{\theta^2 |a|^2 \mathcal{E}_s \mathbf{E}|f(\rvx, \rvz)|^2}{1 - \theta |a|^2 \mathcal{E}_s} - 2 \theta |a| \mathcal{R} \mathbf{E}\left\{e^{j \phi} \bar{f}(\rvx, \rvz) \rvx \right\},
\end{eqnarray}
where $\phi$ is the phase of $a$, and $\mathcal{R}$ denotes the real part of a complex number. By introducing the new variable $\gamma = -\theta |a|^2 \mathcal{E}_s > 0$, we further rewrite $J(a, \theta)$ as
\begin{eqnarray}
J(\gamma, \phi, \theta) = \log(1 + \gamma) - \gamma + \frac{\gamma \theta}{1 + \gamma} \mathbf{E}|f(\rvx, \rvz)|^2 + 2 \sqrt{\frac{-\gamma \theta}{\mathcal{E}_s}} \mathcal{R} \mathbf{E}\left\{e^{j \phi} \bar{f}(\rvx, \rvz) \rvx \right\}.
\end{eqnarray}

Letting the partial derivative $\partial J/\partial \theta$ be zero, we find that the optimal value of $\theta < 0$ should be
\begin{eqnarray}
\sqrt{-\theta_\mathrm{opt}} = \frac{(1 + \gamma) \mathcal{R} \mathbf{E}\left\{e^{j\phi} \bar{f}(\rvx, \rvz) \rvx \right\}}{\mathbf{E}|f(\rvx, \rvz)|^2 \sqrt{\mathcal{E}_s \gamma}}.
\end{eqnarray}
Substituting $\theta_\mathrm{opt}$ into $J(\gamma, \theta)$ followed by some algebraic manipulation, we obtain
\begin{eqnarray}
J(\gamma, \phi, \theta_\mathrm{opt}) = \log(1 + \gamma) - \gamma + \frac{(1 + \gamma)\left[\mathcal{R} \mathbf{E}\left\{ e^{j\phi} \bar{f}(\rvx, \rvz) \rvx \right\}\right]^2}{\mathcal{E}_s \mathbf{E}|f(\rvx, \rvz)|^2}.
\end{eqnarray}
Let us define
\begin{eqnarray}
\Delta(\phi) = \frac{\left[\mathcal{R} \mathbf{E}\left\{ e^{j\phi} \bar{f}(\rvx, \rvz) \rvx \right\}\right]^2}{\mathcal{E}_s \mathbf{E}|f(\rvx, \rvz)|^2},
\end{eqnarray}
and maximize $J(\gamma, \phi, \theta_\mathrm{opt}) = \log(1 + \gamma) - \gamma + (1 + \gamma) \Delta(\phi)$ over $\gamma > 0$. It is straightforward to show that the optimal value of $\gamma$ is $\gamma_\mathrm{opt} = \frac{\Delta(\phi)}{1 - \Delta(\phi)}$, and hence $J(\gamma_\mathrm{opt}, \phi, \theta_\mathrm{opt}) = -\log(1 - \Delta(\phi))$.

It is clear that $J(\gamma_\mathrm{opt}, \phi, \theta_\mathrm{opt})$ is maximized by choosing $\phi = \phi_\mathrm{opt} = - \arctan \mathbf{E}\left\{ \bar{f}(\rvx, \rvz) \rvx \right\}$, which maximizes $\Delta(\phi)$. Denote $\Delta(\phi_\mathrm{opt})$ by $\Delta_\mathrm{opt}$, which is
\begin{eqnarray}
\Delta_\mathrm{opt} = \frac{\left|\mathbf{E}\left\{ \bar{f}(\rvx, \rvz) \rvx \right\}\right|^2}{\mathcal{E}_s \mathbf{E}|f(\rvx, \rvz)|^2}.
\end{eqnarray}

Therefore, the maximum value $J(\gamma_\mathrm{opt}, \phi_\mathrm{opt}, \theta_\mathrm{opt})$, {\it i.e.}, the GMI, is
\begin{eqnarray}
I_\mathrm{GMI} = J(\gamma_\mathrm{opt}, \phi_\mathrm{opt}, \theta_\mathrm{opt}) = \log \left(1 + \frac{\Delta_\mathrm{opt}}{1 - \Delta_\mathrm{opt}}\right) = \log(1 + \mathrm{SNR}_e),
\end{eqnarray}
and the optimal choice of the decoding scaling parameter $a$ is $a_\mathrm{opt} = {\mathbf{E}\left\{ f(\rvx, \rvz) \bar{\rvx} \right\}}/{\mathcal{E}_s}$.

\subsection{Derivation of Eqn. (\ref{eqn:2M-ADC-Delta-1}) and (\ref{eqn:2M-ADC-Delta-2})}
\label{appendix:2M-ADC-Delta-proof}

\begin{eqnarray*}
&&\mathbf{E}[f(\rvx + \rvz)]^2 = 2 \sum_{i = 1}^M \iint_{\alpha_{i - 1} \leq x + z < \alpha_i} r_i^2 p_\rvx(x) p_\rvz(z) dxdz\nonumber\\
&=& 2 \sum_{i = 1}^M r_i^2 \int_{\alpha_{i - 1}}^{\alpha_i} \frac{\exp\left(-\frac{y^2}{2(\mathcal{E}_s + \sigma^2)}\right)}{\sqrt{2\pi (\mathcal{E}_s + \sigma^2)}} dy = 2 \sum_{i = 1}^M r_i^2 \left[Q\left(\frac{\alpha_{i - 1}}{\sqrt{\mathcal{E}_s + \sigma^2}}\right) - Q\left(\frac{\alpha_i}{\sqrt{\mathcal{E}_s + \sigma^2}}\right)\right],\\
&&\mathbf{E}[f(\rvx + \rvz) \rvx] = 2\sum_{i = 1}^M \iint_{\alpha_{i - 1} \leq x + z < \alpha_i} r_i x p_\rvx(x) p_\rvz(z) dxdz\nonumber\\
&=& 2\sum_{i = 1}^M r_i \int_{-\infty}^\infty p_\rvz(z) \left(\int_{\alpha_{i - 1} - z}^{\alpha_i - z} x p_\rvx(x) dx\right) dz\nonumber\\
&=& 2\sum_{i = 1}^M r_i \left[\int_{-\infty}^\infty p_\rvz(z) F(\alpha_{i - 1} - z)dz - \int_{-\infty}^\infty p_\rvz(z) F(\alpha_i - z)dz \right]\nonumber\\
&=& \mathcal{E}_s\sqrt{\frac{2}{\pi(\mathcal{E}_s + \sigma^2)}}\sum_{i = 1}^M r_i \left[e^{-\frac{\alpha_{i - 1}^2}{2(\mathcal{E}_s + \sigma^2)}} - e^{-\frac{\alpha_i^2}{2(\mathcal{E}_s + \sigma^2)}}\right].
\end{eqnarray*}

\subsection{Nearest-Neighbor Decoding for Antipodal Input and Symmetric Output Quantizers}

For a given $2M$-level symmetric quantizer, and for antipodal inputs, we can evaluate the GMI following the result in Section \ref{sec:basic-model-real}. Denote the probability $\mathrm{Pr}[\rvw = r_i | \rvx = \sqrt{\mathcal{E}_s}]$ by $p_i^{(+)}$ and $\mathrm{Pr}[\rvw = -r_i | \rvx = \sqrt{\mathcal{E}_s}]$ by $p_i^{(-)}$; by symmetry, we have $\mathrm{Pr}[\rvw = r_i | \rvx = -\sqrt{\mathcal{E}_s}] = p_i^{(-)}$ and $\mathrm{Pr}[\rvw = -r_i | \rvx = -\sqrt{\mathcal{E}_s}] = p_i^{(+)}$, and $\mathrm{Pr}[\rvw = r_i] = \mathrm{Pr}[\rvw = -r_i] = (p_i^{(+)} + p_i^{(-)})/2$. The GMI thus is
\begin{eqnarray}
I_\mathrm{GMI} = \sup_{t \in \mathbb{R}} \left(t \sqrt{\mathcal{E}_s}\sum_{i = 1}^M (p_i^{(+)} - p_i^{(-)})r_i - \sum_{i = 1}^M (p_i^{(+)} + p_i^{(-)}) \log\cosh(t \sqrt{\mathcal{E}_s} r_i)\right).
\end{eqnarray}
Maximizing GMI with respect to the reconstruction points $\underline{r}$, we have that the optimal $\underline{r}$ satisfies
\begin{eqnarray}
r_i = \frac{1}{t\sqrt{\mathcal{E}_s}} \mathrm{artanh}\left(\frac{p_i^{(+)} - p_i^{(-)}}{p_i^{(+)} + p_i^{(-)}}\right) = \frac{1}{2t\sqrt{\mathcal{E}_s}} \log \frac{p_i^{(+)}}{p_i^{(-)}}, \quad i = 1, \ldots, M,
\end{eqnarray}
and that the GMI further reduces into
\begin{eqnarray}
I_\mathrm{GMI} &=& \sum_{i = 1}^M \left[\frac{p_i^{(+)} - p_i^{(-)}}{2} \log \frac{p_i^{(+)}}{p_i^{(-)}} + (p_i^{(+)} + p_i^{(-)})\log 2 - (p_i^{(+)} + p_i^{(-)}) \log\left(\sqrt{\frac{p_i^{(+)}}{p_i^{(-)}}} + \sqrt{\frac{p_i^{(-)}}{p_i^{(+)}}}\right) \right]\nonumber\\
&=& \log 2 - \sum_{i = 1}^M \left[(p_i^{(+)} + p_i^{(-)}) \log (p_i^{(+)} + p_i^{(-)}) - p_i^{(+)} \log p_i^{(+)} - p_i^{(-)} \log p_i^{(-)}\right] = I(\rvx; \rvw).
\end{eqnarray}
That is, the GMI coincides with the channel input-output mutual information, which is achievable by maximum-likelihood decoding. This seemingly surprising result is in fact reasonable, because there is indeed a nearest-neighbor decoding realization of the maximum-likelihood decoding rule, when the channel input is antipodal and the output quantization is symmetric. Choosing the reconstruction points as $r_i = \log [p_i^{(+)}/p_i^{(-)}]$, $i = 1, \ldots, M$, and denoting $w_k$ by $r_{w_k} \cdot \mathrm{sgn}(w_k)$, we can write the nearest-neighbor decoding metric as
\begin{eqnarray}
\label{eqn:nn-metric-mbit-antipodal}
D(m) &=& \frac{1}{n} \sum_{k = 1}^n \left[\log \frac{p_{r_{w_k}}^{(+)}}{p_{r_{w_k}}^{(-)}} \mathrm{sgn}(w_k) - x_k(m) \right]^2\nonumber\\
&=& \frac{1}{n}\sum_{k = 1}^n \left[\log \frac{p_{r_{w_k}}^{(+)}}{p_{r_{w_k}}^{(-)}}\right]^2 + \mathcal{E}_s - \frac{2}{n}\sum_{k = 1}^n \log \frac{p_{r_{w_k}}^{(+)}}{p_{r_{w_k}}^{(-)}} \mathrm{sgn}(w_k) x_k(m).
\end{eqnarray}
The first two terms in (\ref{eqn:nn-metric-mbit-antipodal}) are independent of the codeword, and thus it suffices to examine
\begin{eqnarray}
D_1(m) = \frac{1}{n}\sum_{k = 1}^n \log \frac{p_{r_{w_k}}^{(+)}}{p_{r_{w_k}}^{(-)}} \mathrm{sgn}(w_k) x_k(m),
\end{eqnarray}
which can be further equivalently deduced into
\begin{eqnarray}
D_2(m) &=& \frac{1}{2n\sqrt{\mathcal{E}_s}} \sum_{k = 1}^n \left[\log \frac{p_{r_{w_k}}^{(+)}}{p_{r_{w_k}}^{(-)}} \mathrm{sgn}(w_k x_k(m)) + \log (p_{r_{w_k}}^{(+)}p_{r_{w_k}}^{(-)})\right]\nonumber\\
&=& \frac{1}{n\sqrt{\mathcal{E}_s}} \sum_{k = 1}^n \log \mathrm{Pr}[w_k|x_k(m)],
\end{eqnarray}
identical to the metric in maximum-likelihood decoding.

\subsection{Super-Nyquist Output Sampling with Antipodal Inputs}

We examine the scenario where the input is antipodal, and where the decoder follows the linearly weighted nearest-neighbor decoding rule:
\begin{eqnarray}
D(m) = \frac{1}{n} \sum_{k = 1}^n \left[\sum_{l = 0}^{L - 1} \beta_l w_{k, l} - x_k(m)\right]^2, \quad m \in \mathcal{M}.
\end{eqnarray}
Following the same line of analysis as that for the Gaussian codebook ensemble, we have, for $m = 1$,
\begin{eqnarray}
\lim_{n \rightarrow \infty} D(1) = \mathbf{E}\left[\sum_{l = 0}^{L - 1} \beta_l \rvw_{0, l} - \rvx_0\right]^2\quad \mbox{a.s.}
\end{eqnarray}
and for any $m \neq 1$,
\begin{eqnarray}
\Lambda(\theta) &=& \lim_{n \rightarrow \infty} \frac{1}{n} \Lambda_n(n\theta)\nonumber\\
&=& \theta \mathbf{E}\left[\left(\sum_{l = 0}^{L - 1} \beta_l \rvw_{0, l}\right)^2\right] + \theta\mathcal{E}_s + \mathbf{E}\left[\log\cosh(2\theta \sqrt{\mathcal{E}_s}\sum_{l = 0}^{L - 1}\beta_l \rvw_{0, l})\right] \quad \mbox{a.s.}
\end{eqnarray}
where $\{\rvw_{0, l}\}_{l = 0}^{L - 1}$ are induced by an infinite sequence of inputs, $\{\rvx_k\}_{k = -\infty}^\infty$. Through some manipulations, we thus obtain the resulting GMI as
\begin{eqnarray}
I_\mathrm{GMI} = \sup_{\underline{\beta}} \left\{\mathbf{E}\left[\rvx_0 \sum_{l = 0}^{L - 1}\beta_l \rvw_{0, l}\right] - \mathbf{E}\left[\log\cosh(\sqrt{\mathcal{E}_s}\sum_{l = 0}^{L - 1}\beta_l \rvw_{0, l})\right]\right\}.
\end{eqnarray}
Consequently, the optimal choice of the weighting coefficients, $\underline{\beta}$, obeys
\begin{eqnarray}
\mathbf{E}\left[\rvw_{0, l}\cdot\tanh\left(\sqrt{\mathcal{E}_s} \sum_{j = 0}^{L - 1}\beta_j \rvw_{0, j}\right)\right] = \mathbf{E}\left[\frac{\rvx_0 \rvw_{0, l}}{\sqrt{\mathcal{E}_s}}\right], \quad l = 0, 1, \ldots, L - 1,
\end{eqnarray}
which constitute an array of transcendental equations.

We further focus on the special case of binary symmetric quantizer $w = \mathrm{sgn}(x + z)$ and $L = 2$. From the symmetry in the setup, we see that $\beta_0 = \beta_1 = \beta$, and we only need to solve a single equation:
\begin{eqnarray}
\label{eqn:solve-beta}
\mathbf{E}[\rvw_{0, 0} \cdot \tanh(\sqrt{\mathcal{E}_s}\beta (\rvw_{0, 0} + \rvw_{0, 1}))] = \frac{1}{\sqrt{\mathcal{E}_s}} \mathbf{E}[\rvx_0 \rvw_{0, 0}].
\end{eqnarray}
For convenience, we denote $\mathrm{Pr}[(\rvw_{0, 0}, \rvw_{0, 1}) = (1, 1)] = \mathrm{Pr}[(\rvw_{0, 0}, \rvw_{0, 1}) = (-1, -1)] = \eta$, $\mathrm{Pr}[(\rvw_{0, 0}, \rvw_{0, 1}) = (1, -1)] = \mathrm{Pr}[(\rvw_{0, 0}, \rvw_{0, 1}) = (-1, 1)] = 1/2 - \eta$, and $\mathrm{Pr}[\rvw_{0, 0} = 1|\rvx_0 = \sqrt{\mathcal{E}_s}] = \kappa$. So (\ref{eqn:solve-beta}) becomes
\begin{eqnarray}
\tanh(2\sqrt{\mathcal{E}_s}\beta) = \frac{2\kappa - 1}{2\eta}, \quad i.e.,\quad \beta = \frac{1}{4\sqrt{\mathcal{E}_s}} \log \frac{2(\eta + \kappa) - 1}{2(\eta - \kappa) + 1}.
\end{eqnarray}

\subsection{Derivation of the GMI in Proposition \ref{prop:gmi-super-nyquist}}
\label{appendix:proof-gmi-super-nyquist}

Denoting the expression in the right-hand side of (\ref{eqn:i-gmi-super-nyquist}) by $J(\underline{\beta}, \theta)$, and enforcing its partial derivatives with respect to $\{\beta_l\}_{l = -L+1}^{L - 1}$ to vanish, we have
\begin{eqnarray}
\frac{\partial J}{\partial \beta_l} &=& 2\theta \mathbf{E}\left[\left(\sum_{u = -L+1}^{L - 1} \beta_u \rvw_{0, u} - \rvx_0\right) \rvw_{0, l}\right] - \frac{2\theta}{1 - 2\theta \mathcal{E}_s} \mathbf{E} \left[\left(\sum_{u = -L+1}^{L - 1} \beta_u \rvw_{0, u}\right) \rvw_{0, l}\right] = 0\nonumber\\
\Rightarrow &&\sum_{u = -L+1}^{L - 1} \beta_u \mathbf{E}[\rvw_{0, u} \rvw_{0, l}] = \left(1 - \frac{1}{2\theta \mathcal{E}_s}\right) \mathbf{E}[\rvx_0 \rvw_{0, l}],
\end{eqnarray}
for $l = -L+1, \ldots, L - 1$. Summarizing these $2L - 1$ equations, we can write them collectively as
\begin{eqnarray}
\mathbf{\Omega} \underline{\beta} = \left(1 - \frac{1}{2\theta \mathcal{E}_s}\right) \underline{b},
\end{eqnarray}
where $\mathbf{\Omega}$ is a $(2L-1) \times (2L-1)$ matrix with its $(u, l)$-element being $\mathbf{E}[\rvw_{0, u} \rvw_{0, l}]$, and $\underline{b}$ is a $(2L-1)$-dimensional vector with its $l$-element being $\mathbf{E}[\rvx_0 \rvw_{0, l}]$. Hence we have
\begin{eqnarray}
\label{eqn:optimal-beta-super-nyquist}
\underline{\beta} = \left(1 - \frac{1}{2\theta \mathcal{E}_s}\right) \mathbf{\Omega}^{-1} \underline{b}.
\end{eqnarray}
Substituting (\ref{eqn:optimal-beta-super-nyquist}) into $J(\underline{\beta}, \theta)$, we get
\begin{eqnarray}
\label{eqn:J-opt-beta}
J(\underline{\beta}, \theta) &=& \frac{2\theta^2 \mathcal{E}_s}{2\theta \mathcal{E}_s - 1} \sum_{l = -L+1}^{L - 1} \sum_{u = -L+1}^{L - 1} \beta_l \beta_u \Omega_{u, l} + \theta \mathcal{E}_s - 2\theta \sum_{l = -L+1}^{L - 1} \beta_l b_l + \frac{1}{2}\log(1 - 2\theta \mathcal{E}_s)\nonumber\\
&=& \theta \mathcal{E}_s + \left(\frac{1}{2\mathcal{E}_s} - \theta\right) \underline{b}^T \mathbf{\Omega}^{-1} \underline{b} + \frac{1}{2}\log(1 - 2\theta \mathcal{E}_s).
\end{eqnarray}
From (\ref{eqn:J-opt-beta}), we maximize $J(\underline{\beta}, \theta)$ by letting
\begin{eqnarray}
1 - 2\theta\mathcal{E}_s = \frac{\mathcal{E}_s}{\mathcal{E}_s - \underline{b}^T \mathbf{\Omega}^{-1} \underline{b}},
\end{eqnarray}
and the maximum value of $J(\underline{\beta}, \theta)$, {\it i.e.}, the GMI, is
\begin{eqnarray}
I_\mathrm{GMI} = \frac{1}{2}\log\left(1 + \frac{\underline{b}^T \mathbf{\Omega}^{-1} \underline{b}/\mathcal{E}_s}{1 - \underline{b}^T \mathbf{\Omega}^{-1} \underline{b}/\mathcal{E}_s}\right).
\end{eqnarray}

\end{document}

%% file: rv_defs1.tex
\DeclareMathAlphabet{\eurm}{U}{eur}{m}{n}
\DeclareMathAlphabet{\mathbsf}{OT1}{cmss}{bx}{n}
\DeclareMathAlphabet{\mathssf}{OT1}{cmss}{m}{sl}
\DeclareMathAlphabet{\mathcsf}{OT1}{cmss}{sbc}{n}

\newcommand{\randomvalue}[1]{\eurm{\uppercase{#1}}}

\DeclareSymbolFont{bsfletters}{OT1}{cmss}{bx}{n}  
\DeclareSymbolFont{ssfletters}{OT1}{cmss}{m}{n}
\DeclareMathSymbol{\bsfGamma}{0}{bsfletters}{'000}
\DeclareMathSymbol{\ssfGamma}{0}{ssfletters}{'000}
\DeclareMathSymbol{\bsfDelta}{0}{bsfletters}{'001}
\DeclareMathSymbol{\ssfDelta}{0}{ssfletters}{'001}
\DeclareMathSymbol{\bsfTheta}{0}{bsfletters}{'002}
\DeclareMathSymbol{\ssfTheta}{0}{ssfletters}{'002}
\DeclareMathSymbol{\bsfLambda}{0}{bsfletters}{'003}
\DeclareMathSymbol{\ssfLambda}{0}{ssfletters}{'003}
\DeclareMathSymbol{\bsfXi}{0}{bsfletters}{'004}
\DeclareMathSymbol{\ssfXi}{0}{ssfletters}{'004}
\DeclareMathSymbol{\bsfPi}{0}{bsfletters}{'005}
\DeclareMathSymbol{\ssfPi}{0}{ssfletters}{'005}
\DeclareMathSymbol{\bsfSigma}{0}{bsfletters}{'006}
\DeclareMathSymbol{\ssfSigma}{0}{ssfletters}{'006}
\DeclareMathSymbol{\bsfUpsilon}{0}{bsfletters}{'007}
\DeclareMathSymbol{\ssfUpsilon}{0}{ssfletters}{'007}
\DeclareMathSymbol{\bsfPhi}{0}{bsfletters}{'010}
\DeclareMathSymbol{\ssfPhi}{0}{ssfletters}{'010}
\DeclareMathSymbol{\bsfPsi}{0}{bsfletters}{'011}
\DeclareMathSymbol{\ssfPsi}{0}{ssfletters}{'011}
\DeclareMathSymbol{\bsfOmega}{0}{bsfletters}{'012}
\DeclareMathSymbol{\ssfOmega}{0}{ssfletters}{'012}

\newcommand{\rvm}{{\randomvalue{m}}}	

\newcommand{\rvv}{{\randomvalue{v}}}	

\newcommand{\rvw}{{\randomvalue{w}}}	

\newcommand{\rvx}{{\randomvalue{x}}}	

\newcommand{\rvy}{{\randomvalue{y}}}	

\newcommand{\rvz}{{\randomvalue{z}}}	

